\documentclass[]{aa}
\usepackage{graphics}

\title{Unstable quasi g-modes in rotating main-sequence stars}

\author{G.J. Savonije}

\institute{ Astronomical Institute `Anton Pannekoek', University of
  Amsterdam, Kruislaan 403, 1098 SJ Amsterdam} 
\date{received ?? ; accepted ??}

\begin{document}
\abstract{This paper studies the oscillatory stability of uniformly rotating main-sequence stars of mass  3-8 $\mathrm{M}_\odot$ by solving the linearized non-adiabatic, non-radial oscillation equations with a forcing term and searching for resonant response to a complex forcing frequency. By using the traditional approximation the solution of the forced oscillation equations becomes separable, whereby the energy equation is made separable by approximation.
It is found that the $\kappa$-mechanism in rotating B-stars can  destabilize not only gravity- or pressure modes, but also a branch of low frequency retrograde (in corotating frame) oscillations in between the retrograde g-modes and toroidal r-modes. These unstable quasi g-modes (or `q-modes') hardly exhibit rotational confinement to the equatorial regions of the star, while the oscillations are always prograde in the observer's frame, all in contrast to g-modes. The unstable q-modes occur in a few narrow period bands (defined by their azimuthal index m), and seem to fit the oscillation  spectra  observed in SPB stars rather well. The unstable q-mode oscillation spectrum of a very rapidly rotating 8$M_\odot$ star appears similar to that of the well studied Be-star $\mu$ Cen. The unstable q-modes thus seem far better in explaining the observed oscillation spectra in SPB-stars and Be-stars than normal g-modes.
  }
\maketitle

\section{Introduction}
With the introduction of new opacity data by Iglesias et al. (\cite{iglesias92}) and a later update by Iglesias \& Rogers (\cite{iglesias96}) it was suggested that the origin of $\beta$ Cephei pulsations in massive main-sequence stars and of the slowly pulsating B (SPB) stars (Waelkens \cite{waelkens91}) of lower mass is caused by the classical opacity valve mechanism (Eddington \cite{eddington}) operating on non radial g-mode or p-mode oscillations. Early stability analyses with these new opacity results in non-rotating stars have been published by e.g. Kiriakidis et al. (\cite{kiria92}), Gautschy and Saio (\cite{gautschy93}) and Dziembowski et al. (\cite{dziembowski93}). 

Observations by Balona (\cite{balona94}) and Balona and Koen (\cite{balonaK94}) showed an unexplainable lack of SPB stars in two open clusters which do contain many pulsating $\beta$ Cephei stars. It was suggested that the lack of SPB stars in these clusters might be correlated with the rapid rotation of the cluster stars. Ushomirsky and Bildsten (\cite{usho98}), in an attempt to explain Balona and Koen's observations in terms of rotational stabilization of g-modes, applied the traditional approximation combined with a quasi-adiabatic stability analysis to study the effect of rotation on the $\kappa$ instability mechanism. They did not find a decisive solution: rotation seemed indeed to stabilize some pulsation modes but to destabilize other g-modes. This work was followed up by Lee (\cite{lee01}) who used a fully non-adiabatic analysis with a truncated expansion of spherical harmonics to approximate the low-frequency g-mode oscillations in rotating stars. Lee found some rotational stabilization caused by the Coriolis coupling with higher degree spherical harmonic components, especially for retrograde g-modes in the inertial regime, but this seems not sufficient to explain the lack of SPB stars in clusters with rapidly rotating B-stars. Townsend (\cite{town03}), also applying the traditional approximation, tried to explain the lack of observed rapidly rotating SPB stars by the well known effect of the Coriolis force on g-mode pulsations: the confinement of
the oscillation to the equatorial regions ($|\mu|< \bar{\sigma}/(2 \Omega_s)$) of the star, where $\mu=\cos{\vartheta}$ and $\bar{\sigma}$ is the oscillation frequency in the corotating frame (e.g. Savonije et al. \cite{sav95}). The lack of rapidly rotating SPB stars would then be caused by a selection effect, not by rotational stabilization of g-modes. However, more recent observations indicate that the two open clusters studied by Balona and Koen may after all contain a few SPB stars (Stankov et al. \cite{stank02}), so that the problem may be non-existent.

There is, however, another complication: the existence of unstable rotational `q-modes' in rotating main-sequence stars whose instability has been hitherto neglected. These q-modes exhibit hardly any rotational confinement  towards the equatorial region, in contrast to normal g-modes. In this paper we will study the stability of these quasi g-modes in comparison with normal (retrograde) g-modes and check how well unstable q-modes can explain the oscillation spectra of rotating B-stars.

\section{Basic oscillation equations}
We consider  uniformly rotating main sequence stars with mass
$M_\mathrm{s}$ between 3-8 $\mathrm{M}_\odot$ and denote their radius by $R_\mathrm{s}$.
We wish to study the oscillatory stability of these
uniformly rotating B-stars by subjecting them to a perturbing periodic force to determine the resonant behaviour.   
We assume the star's angular velocity of rotation $\vec{\Omega}_\mathrm{s}$ to be
much smaller than its break-up speed, i.e.
$(\Omega_\mathrm{s}/\Omega_\mathrm{c})^2\ll 1$, with
$\Omega_\mathrm{c}^2=GM_\mathrm{s}/R_\mathrm{s}^3$, so that effects of
centrifugal distortion ($\propto \Omega_\mathrm{s}^2$) may be
neglected to a first approximation.
The Coriolis acceleration is proportional to
$\Omega_\mathrm{s}$ and we consider its effect on the induced oscillatory
motions in the star. We use nonrotating spherical coordinates
($r,\vartheta,\varphi)$, with the origin at the star's centre,
whereby $\vartheta=0 $ corresponds to its rotation axis.

Let us denote the displacement vector in the star by $\vec{\xi}$ and
perturbed Eulerian quantities like pressure $P'$, density $\rho'$,
temperature $T'$ and energy flux $\vec{F'}$ with a prime. The
linearized hydrodynamic equations governing the non-adiabatic response
of the uniformly rotating star to the perturbing potential
$\Phi_\mathrm{T}$ may then be written as 
\begin{eqnarray}
&&\left[ \left(\frac{\partial}{\partial t} + \Omega_\mathrm{s} \frac{\partial }
    {\partial\varphi}\right)
  v'_i\right] \vec{e}_i+ 2 \vec{\Omega}_\mathrm{s} \times \vec{v'}=
\nonumber\\
  &&\quad -\frac{1}{\rho}
  \nabla P' + \frac{\rho'}{\rho^2} \nabla P - \nabla \Phi_\mathrm{T},
  \label{eqmot}\label{eq:1}
\end{eqnarray}
\begin{equation}
  \left(\frac{\partial}{\partial t} + \Omega_\mathrm{s} 
    \frac{\partial }{\partial\varphi}\right) \rho' +
  \nabla\cdot\left(\rho \vec{v'} \right) =0, \label{eqcont}
\end{equation}
\begin{equation} 
  \left(\frac{\partial }{\partial t} + \Omega_\mathrm{s}
    \frac{\partial }{\partial \varphi}\right) \left[ S' +
    \vec{\xi}\cdot \nabla S \right]=-\frac{1}{\rho T} \nabla\cdot\vec{F'}, \label{eqe}
\end{equation}
\begin{equation}
  \frac{\vec{F'}}{F}=\left(\frac{\mathrm{d}T}{\mathrm{d}r}\right)^{-1} \left[ \left(\frac{3
        T'}{T} -\frac{\rho'}{\rho} -\frac{\kappa'}{\kappa} \right) \nabla
    T + \nabla T' \right] \label{eqf} 
\end{equation}
where $\kappa$ denotes the opacity of stellar material and $S$ its
specific entropy. These perturbation equations represent,
respectively, conservation of momentum, conservation of mass and
conservation of energy, while the last equation describes the
radiative diffusion of the perturbed energy flux. For simplicity we
adopt the Cowling (1941) approximation, i.e. we neglect perturbations to the 
gravitational potential caused by the star's oscillatory distortion. For the higher radial order oscillation modes studied here, this approximation should be adequate.
As usual, we also neglect perturbations of the nuclear energy
sources (not important) and of convection (frozen convective flux approximation).
In main-sequence (MS)stars more massive than about 6 $M_\odot$ in which the opacity Z-bump driving region is convectively unstable this latter approximation introduces some uncertainty with respect to the detected oscillatory instabilities.

For the periodic forcing we apply a force equal to the (real part of) the gradient of
\begin{equation}
  \Phi_\mathrm{T}(r,\vartheta,\varphi,t)= f r^l \, P^m_l (\mu) \,e^{ \mathrm{i} 
(\sigma  t -m \varphi)} \label{eqpot}
\end{equation}
where $\sigma$ is the forcing frequency in the inertial frame,
$\mu=\cos\vartheta$, $P^m_l(\mu)$ is an associated Legendre polynomial 
and $ f$ is an arbitrary constant. Since we are interested in periodic solutions of the
forced oscillations we assume the perturbations have the same time and $\varphi$ dependence as
the forcing (although with a certain phase lag due to radiative damping).
After factoring out the known time and $\varphi$-dependence, Eqs.~(\ref{eqmot}--\ref{eqf})
form a 2D problem in the ($r,\vartheta$) meridional plane of the perturbed star.
The 2D problem requires a large amount of computer memory and CPU time to solve, see e.g. Savonije et al. (\cite{sav95}). However, it is possible to obtain solutions adequate for initial analyses of the stability of rotating stars by applying the so called `traditional approximation'.

\section{The traditional approximation}
For a rotating star the solutions of Eqs.~(\ref{eqmot}--\ref{eqf})  are no longer separable into $r$-,$\vartheta$- and $\varphi$- factors and cannot be described by a single spherical harmonic $(l, m)$ due to the action of the Coriolis force on oscillating stellar matter. However, for a uniformly rotating star in the traditional approximation separability is retained by neglecting the $\vartheta$- component of the angular velocity vector (e.g. Unno et al. \cite{unno89}). This is done because the radial motions are expected to be small in the stably stratified layers of the star, especially for low-frequency modes (Savonije et al. \cite{sav95}).
In this way the problem is reduced to two 1-dimensional problems (in $\vartheta$ and $r$) with a prescribed azimuthal harmonic dependence $e^{i m \varphi}$ since we assume the unperturbed star to be axisymmetric. 
When only the radial component of $\vec{\Omega}_\mathrm{s}$ is retained, the
$\vartheta$- and $\varphi$-components of the equation of motion can be
written: 
\begin{equation}
  -{\bar{\sigma}}^2\xi_{\vartheta} -2\mathrm{i}\Omega_\mathrm{s}{\bar{\sigma}}\cos\vartheta
  \xi_{\varphi}=
  -\frac{1}{r\rho}\frac{\partial P'}{\partial \vartheta} -\frac{1}{r}\frac{\partial
    \Phi_\mathrm{T}}{\partial \vartheta},\label{CRAP}
\end{equation}
\begin{equation}
  -\bar{\sigma}^2\xi_{\varphi}
  +2\mathrm{i}\Omega_\mathrm{s}\bar{\sigma}\cos\vartheta \xi_{\vartheta}=
  \frac{\mathrm{i}m}{r\rho\sin\vartheta } P'
  +\frac{\mathrm{i}m}{r\sin\vartheta} \Phi_\mathrm{T}.\label{CRAP1}
\end{equation}
We have expressed the velocity perturbations in terms of the
displacement vector by the relation $\vec{v'}=\mathrm{i} \bar{\sigma}
\vec{\xi}$, with $\bar{\sigma}=\sigma-m\, \Omega_\mathrm{s}$ the
oscillation frequency in the corotating stellar frame (negative if the oscillation is retrograde, i.e. propagates in the direction counter to the stellar rotation). 
We will only consider positive $m$-values in this paper, except when explicitely stated
otherwise (in Tables~\ref{compare_1}-\ref{compare_2}).

\subsection{Separation of variables}
A separation of variables can be performed by writing, see Papaloizou \& Savonije (\cite{pap97}): 
\[ 
\xi_{\vartheta}=\sum_{n=0}^{\infty}\mathcal{F}_n(\vartheta)\,
D_n(r) \mbox{\hspace{1cm}}
\xi_{\varphi}=\sum_{n=0}^{\infty}\mathcal{G}_n(\vartheta)\, D_n(r)
\]
\begin{equation}
P' =\sum_{n=0}^{\infty} \mathcal{X}_n(\vartheta) \,  W_n(r),
\end{equation}
with $\xi_r$, $T'$ and $\rho'$ having expansions of the same form as that for
$P'$. Here we leave out the known factor $e^{i m \varphi}$ for the azimuthal variation of the perturbed quantities, while $\mathcal{F}_n,\mathcal{G}_n$ and $\mathcal{X}_n$ are functions of $\vartheta$ chosen to obey the relations
\[ 
-\bar{\sigma}^2\, \mathcal{F}_n
-2\mathrm{i}\Omega_\mathrm{s}{\bar{\sigma}}\cos\vartheta
\, \mathcal{G}_n= -{\frac{\mathrm{d} \mathcal{X}_n}{\mathrm{d}\vartheta}} 
\] 
and 
\[
-{\bar{\sigma}}^2 \, \mathcal{G}_n
+2\mathrm{i}\Omega_\mathrm{s}\bar{\sigma}\cos\vartheta \, \mathcal{F}_n=
\frac{\mathrm{i}m}{\sin\vartheta}\, \mathcal{X}_n.
\]

We obtain an equation for $\mathcal{X}_n(\vartheta)$ by imposing the constraint
\[ 
\frac{1}{\sin\vartheta}\frac{\mathrm{d}\left( \sin\vartheta\mathcal{F}_n\right)}
 {\mathrm{d} \vartheta} -\mathrm{i} m\frac{\mathcal{G}_n}{\sin\vartheta}
=-\frac{\lambda_n}{{\bar{\sigma}}^2} \mathcal{X}_n,
\] 
where $\lambda_n$ is a constant. Then $\mathcal{X}_n$ must satisfy the second-order
equation obtained from
\begin{equation}
  \frac{1}{\sin\vartheta}\frac{\mathrm{d} \mathcal{Q}_n}{\mathrm{d}\vartheta} +
\frac{m x \cos\vartheta}{\sin^2\vartheta}\mathcal{Q}_n
=\mathcal{X}_n\left(\frac{m^2}{\sin^2\vartheta}-\lambda_n\right)\label{lamb1}
\end{equation}
with
\begin{equation}
  \mathcal{Q}_n=\frac{\sin\vartheta}{(1-x^2\cos^2\vartheta)}\left( \frac{\mathrm{d} \mathcal{X}_n}  {\mathrm{d}\vartheta} -\frac{m x \cos\vartheta}{\sin\vartheta} \mathcal{X}_n\right)\label{lamb2}
\end{equation}
whereby $\mathcal{X}_n(\vartheta)$ is an eigenfuction with $\lambda_n$ the
associated eigenvalue. The solution depends on the rotation parameter
$x=2\Omega_\mathrm{s}/{\bar{\sigma}}$. 
For $\Omega_\mathrm{s}=0$ the functions $\mathcal{X}_n(\vartheta)$ become the associated 
Legendre functions $P^{m}_{m+n}(\cos\vartheta)$ with corresponding eigenvalues 
$\lambda_n=(m+n)~(m+n+1).$ Normal modes of the rotating star correspond to normal
)modes of the non-rotating star with $(m+n)~(m+n+1)$ replaced by any
permissible $\lambda_n.$

Different $\mathcal{X}_n(\vartheta)$ are orthogonal on integration with respect
to $\mu=\cos\vartheta$ over the interval $(-1,1)$.
If the perturbing potential is expanded in terms of the $\mathcal{X}_n$ such
that
\begin{equation}
  \Phi_\mathrm{T}(r,\vartheta) =\sum_{n=1}^{\infty} \Psi_n(r)\, \mathcal{X}_n(\vartheta), \label{potexp}
\end{equation} 
we find from~(\ref{CRAP}) and~(\ref{CRAP1}) that
\begin{equation}
  D_n(r)=\frac{1}{r\rho}W_n(r)+\frac{1}{r} \Psi_n(r).
\end{equation}
This equation is exactly the same as in the non-rotating case. The
same is true for the equation of continuity, except that
$(m+n)~(m+n+1)$ is replaced by $\lambda_n$. In this way the adiabatic stellar response will consist of a superposition of responses appropriate to non-rotating stars with $(m+n)~(m+n+1)$ replaced by $\lambda_n$ obtained from equations (\ref{lamb1}-\ref{lamb2}). However, we are interested in the oscillatory stability of rotating stars and for that we need to include non-adiabatic effects, i.e. we have to consider the energy equation (\ref{eqe}). We can write the divergence term on its right hand side as:
\[
\nabla \cdot \frac{\vec{F'}}{F_r}=
 \frac{\partial {\left(r^2 \frac{F'_r}{F_r}\right)}} { r^2 \partial r} + \frac{1}{r \sin{\vartheta}} \frac{\partial \left(\sin{\vartheta} \frac{F'_\vartheta}{F_r} \right)} {\partial \vartheta} - \frac{\mathrm{i} m}{r \sin{\vartheta}} \left(\frac{F'_\varphi}{F_r}\right)
\]
By adopting $T'(r,\vartheta)=\sum_n T'_n(r) \, \mathcal{X}_n(\vartheta)$ and substitution in the flux equation (\ref{eqf}) the radial part of the above divergence can be written in the required separated form $\mathcal{D}_r (r, \vartheta)=H(r) \,\mathcal{X}_n(\vartheta)$.  However, the two angular terms cannot be expressed in this way. For one can write, after applying equation (\ref{eqf}) and equations (\ref{lamb1}-\ref{lamb2}),  the angular part  of the above divergence as:
\[ \mathcal{D}_a(r,\vartheta)=\frac{1}{r^2}\left(\frac{\mathrm{d} \log{T}}{\mathrm{d} r}\right)^{-1} \frac{T'(r)}{T(r)}\,\, \mathcal{Y}_n (\vartheta) \,\mathcal{X}_n(\vartheta) \]
with
\[
 \mathcal{Y}_n(\vartheta)=2 \mu \, x^2\, \frac{\mathcal{Q}_n(\mu)}{\mathcal{X}_n(\mu)} + \left(x^2 \, \mu^2 \,  - m\, x - \lambda_n \right) \]
which shows that the energy equation renders the system of equations inseparable, unless $x \rightarrow 0$. Nevertheless, we wish to apply the traditional approximation and take advantage of its simple treatment of rotational effects. Therefore we will approximate the energy equation by a separable equation  by averaging the angular part of the divergence over $\vartheta$ and using  the averaged value $\epsilon_n$ for $\mathcal{Y}_n$:

\[ \epsilon_n=\frac{\int \mathcal{Y}_n \,\mathcal{X}_n\, \mathrm{d} \mu}{\int \mathcal{X}_n\, \mathrm{d}\mu}  \]
so that
\begin{equation}
\mathcal{D}_a(r,\vartheta)\simeq \frac{1}{r^2}\left(\frac{\mathrm{d} \log{T}}{\mathrm{d} r}\right)^{-1} \frac{T'(r)}{T(r)}\,\, \epsilon_n \,\mathcal{X}_n(\vartheta)  \label{D_a}
\end{equation}
For small values of the rotation-parameter  $x \rightarrow 0$ the exact result $\epsilon_n \rightarrow -\lambda_n=-(m+n)~(m+n+1)$ for non-rotating stars is approached. 

Expression (\ref{D_a}) takes  the radiative diffusion in the horizontal direction at least qualitatively into account.  Below we will check our results by comparing with other linear stability calculations of rotating stars.

\subsection{The radial part of the oscillation equations}
\label{sosceq}
Once we have solved the above angular eigenvalue problem for $\lambda_n$ we can factor out the common, now known, factor $\mathcal{X}_n(\vartheta)\, e^{\mathrm{i}(\sigma \,t-m\, \varphi)}$ from all perturbed quantities. The linearized equations describing the non-adiabatic forced oscillations, for the $n$-th  component in expansion~(\ref{potexp}), then form a 1-dimensional (radial) problem.
The remaining radial part of the oscillation equations can  be expressed as:
\begin{equation}
  \bar{\sigma}^2 \rho \xi_r-\frac{\mathrm{d}P'}{\mathrm{d}r}+\frac{\mathrm{d}P}{\mathrm{d}r} \left(\frac{\rho'}{\rho}\right)= 
  l\, \rho\, f\,  r^{\left(l-1\right)}, \label{eq1} 
\end{equation}
\begin{equation}
  \frac{1}{\rho r^2} \frac{\mathrm{d}\left(\rho r^2 \xi_r\right)}{\mathrm{d}r}= -\frac{\rho'}{\rho} + 
  \frac{\lambda_n}{\bar{\sigma}^2 r^2} \left[\frac{P}{\rho}\left(\frac{P'}{P}\right) + 
    f\, r^l\, \right],  \label{eq2}
\end{equation}
\begin{equation}
  \frac{P'}{P} - \Gamma_1 \frac{\rho'}{\rho} + \mathcal{A}\, \xi_r =
  \mathrm{i} \eta \left[ -\frac{1}{F r^2} \frac{\mathrm{d}(r^2 F'_r)}{\mathrm{d}r} + \epsilon_n     \frac{\Lambda}{r^2}\left(\frac{T'}{T}\right) \right], \label{eq3}
\end{equation}
\begin{equation}
  \frac{F'_r}{F}=\Lambda \frac{\mathrm{d}}{\mathrm{d}r} \left(\frac{T'}{T}\right)
  + (4 - \kappa_T) \frac{T'}{T} -(1 + \kappa_{\rho}) \frac{\rho'}{\rho} \label{eq4}
\end{equation}
where $\Gamma$'s represent Chandrasekhar's adiabatic coefficients,
\[
\Lambda= \left(\frac{\mathrm{d}\log T}{\mathrm{d}r}\right)^{-1}; \mbox{\hspace{.5cm}}
\mathcal{A}=\frac{\mathrm{d}\log P}{\mathrm{d}r}- \Gamma_1 \frac{\mathrm{d}\log \rho}
{\mathrm{d}r},
\]
the opacity derivatives are given by:
\[ 
\kappa_T=\left(\frac{\partial\log \kappa}{\partial \log T}\right)_{\rho};
\mbox{\hspace{.3cm}} \kappa_{\rho}=\left(\frac{\partial\log \kappa}{\partial \log
    \rho}\right)_T ;
\]

$\eta$ is a characteristic radiative diffusion length:
\[ 
\eta=-(\Gamma_3-1) F/\left(\bar{\sigma} P\right), \mbox{\,\,\, with
  \hspace{0.3cm}} F=- \frac{4 a c T^3}{3 \kappa \rho} \frac{\mathrm{d}T}{\mathrm{d}r}
\]
being the unperturbed (radial) radiative energy flux, $\epsilon_n$ is related to the horizontal radiative diffusion and is defined by (\ref{D_a}). All other constants
have their usual meaning.   Note that the radiative diffusion
introduces a factor $\mathrm{i}$, so that the radial parts of the perturbations are
complex-valued. This expresses the induced phase-lags with respect to the
external forcing caused by non-adiabatic effects due to radiative diffusion.

The oscillation equations are complemented by the linearized equation of state
\begin{equation}
  \frac{P'}{P}=\left(\frac{\partial\log P}{\partial \log T}\right) \frac{T'}{T}+ 
  \left(\frac{\partial\log P}{\partial \log \rho}\right) \frac{\rho'}{\rho}+
  \left(\frac{\partial\log P}{\partial \log \mu_a}\right) \frac{\mu'_a}{\mu_a} 
\end{equation} 
with 
\[ \frac{\mu'_a}{\mu_a}=-\frac{\mathrm{d} \log{\mu_a}}{\mathrm{d} r}\, \xi_r \]
where $\mu_a$ is the mean atomic weight of the stellar gas. 

Finally, we prescribe the usual boundary conditions  $\xi_r=F'=0$ at the stellar centre and require
that the Lagrangian perturbations at the stellar surface obey:
$\delta P=0$ and $\delta F/F=4\, \delta T/T$ (Stefan's law).

\subsection{Stellar input models for the oscillation code}
We constructed the unperturbed stellar models for the main-sequence stars
with a recent version (Pols et al. \cite{pols95}) of the stellar
evolution code developed by Eggleton (\cite{egg72}).  The models
represent spherical main-sequence stars  with masses $4-8 M_\odot$ and
chemical composition given by various values of the central hydrogen
abundance $X_\mathrm{c}$ and $Z=0.02$ or $Z=0.03$.  These models were constructed by
using the OPAL opacities (Iglesias \& Rogers \cite{iglesias96}).  Table~\ref{inputm} lists the effective temperatures of the various stellar input models characterized by their mass, core hydrogen mass fraction $X_c$ and metal content $Z$,  used in our calculations.
\begin{table}[htbp]
  \caption{($^{10} \log(L/L_\odot), ^{10} \log T_\mathrm{eff})$ of the input models;
an asterisk above stellar mass indicates models with Z=0.03 instead of Z=0.02}
\[
   \begin{array}{|c|c|c|c|}  \hline 
\mbox{Model} &  X_c=0.6 & X_c=0.4 & X_c=0.2 \\
\hline
\hline
3 M_\odot & (1.932, 4.069) & (2.025,4.037) & \\
\hline
4 M_\odot  &  (2.411,4.153) & (2.517,4.124) & (2.597,4.0790 \\
\hline
6 M_\odot  & &  (3.173,4.238) & (3.267,4.199)  \\
\hline
6^* M_\odot  & & (3.112,4.210) & (3.205,4.167)  \\
\hline
8^* M_\odot  & & (3.562,4.288) & (3.662,4.247) \\
\hline
  \end{array} 
\] \label{inputm}
\end{table}

\section{Solution method}
\subsection{Determination of the eigenvalue $\lambda_n(x)$}
For a given stellar rotation frequency $\Omega_\mathrm{s}$, forcing frequency in corotating frame $\bar{\sigma}$, azimuthal index $m$ and angular order $n$, the eigenvalues $\lambda_n$ can be determined by numerically integrating the two first order differential Eqs.~(\ref{lamb1}) and~(\ref{lamb2}).
We have used a shooting method with fourth order Runge-Kutta
integration with variable stepsize (Press et al. \cite{press92}) to obtain numerical solutions.  
It is convenient to rewrite equations (\ref{lamb1}-\ref{lamb2}) as
\begin{equation}
  \frac{\mathrm{d}\mathcal{X}_n}{\mathrm{d}\mu}=-\frac{m x\, \mu}{1-\mu^2}\,\mathcal{X}_n -\left(\frac{1-x^2 \mu^2}
    {1-\mu^2} \right) \,\mathcal{Q}_n, \label{Lamb1} 
\end{equation}
\begin{equation}
  \frac{\mathrm{d}\mathcal{Q}_n}{\mathrm{d}\mu}=-\left(\frac{m^2}{1 -\mu^2} -\lambda_n\right)\, 
\mathcal{X}_n +
  \frac{m x \mu}{1 -\mu^2} \,\mathcal{Q}_n \label{Lamb2} 
\end{equation}
where the factor $x=2 \Omega_\mathrm{s}/\bar{\sigma}$ expresses the
importance of stellar rotation.

To enable integration away from $\mu=1$ it is convenient to write $
\mathcal{X}_n(\mu)=(1 -\mu^2)^{\frac{m}{2}} Y_n(\mu) $ and expand $Y_n(\mu)$ in
a power series which can be substituted in Eqs.~(\ref{Lamb1}) and~(\ref{Lamb2}) to determine
the coefficients. 
$\mathcal{Q}_n$ can then be expressed in terms of $Y_n$ as
\[ 
\mathcal{Q}_n=\frac{(1-\mu^2)^{\frac{m}{2}}} {1-x^2 \mu^2} \, \left[ m
  \mu (1-x)\, Y_n-(1-\mu^2)\, \frac{\mathrm{d}Y_n}{\mathrm{d}\mu} \right].
\] 
For the even $n$ (including $n$=0) solutions the boundary conditions 
are $\mathcal{X}_n=0$ at $\mu=1$ and $\mathcal{Q}_n=0$ at $\mu=0$, while the odd solutions
are defined by  $\mathcal{X}_n=0$ at $\mu=0$.
We can now integrate equations (\ref{Lamb1})-(\ref{Lamb2}) from 
$\mu=1-\delta$ (with $\delta=10^{-4}$) to
$\mu=0$ with an estimated value for the eigenvalue $\lambda_n$. We
iterate by adjusting $\lambda_n$ until the integrated value for either
$\mathcal{Q}_n$ (even solutions), or $\mathcal{X}_n$ (odd solutions) is sufficiently close to zero 
for $\mu=0$.
\begin{figure}[htbp]
  \includegraphics{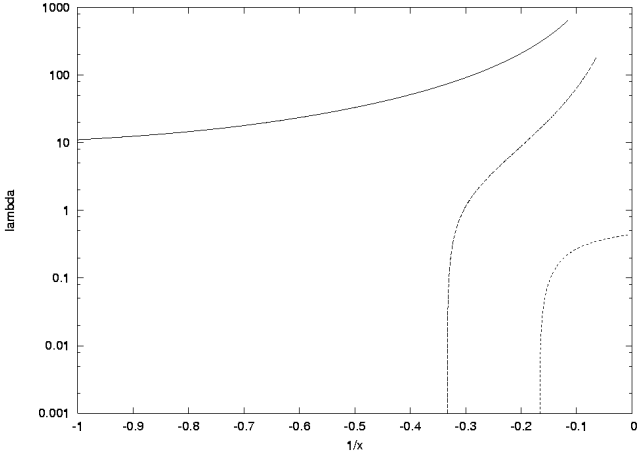}
  \caption{The eigenvalue $\lambda_n$ versus $\bar{\sigma}/(2 \Omega_s)$ for $m$=2 modes. From top to bottom right:  full line: g-modes with $n=0$; dashed line: q-modes with $n=-1$; dotted line: q-modes with $n=-2$. The $n=-1$ branch of quasi g-modes lies between the normal g-modes and the quasi-toroidal modes (with $\lambda_{-2}$ remaining small as $|1/x|\rightarrow 0)$. The eigenfunction $\mathcal{X}_n(\vartheta)$ corresponding to $n=0$ and $n=-2$ has no nodes, that of $n=-1$ has a node at the equator.}
  \label{lambda} 
\end{figure}

\subsection{Eigenvalues $\lambda_{n m}$ and mode classification ($n$,$m$)}
Eigenvalues $\lambda_{n m}$ of equations (\ref{lamb1}-\ref{lamb2}) (or its equivalent) have recently been calculated by  Bildsten et al. (\cite{bild96}), Papaloizou \& Savonije (\cite{pap97}), Lee \& Saio (\cite{lee97}) and Savonije \& Witte (\cite{sav02}).
It appears that in the traditional approximation the angular eigenvalue $\lambda_{n m}$ can take both positive and negative values, including 0. Here we will follow Lee \& Saio (1997, see their Fig.~1) and allow $n$ to assume positive and negative values (or 0) to discriminate between these cases. The positive values $n=0,1,2,3$,... correspond to gravity (g)-modes (or p-modes) for which $\lambda>0$ for any $x$-value, whereby even values of $n$ (or 0) correspond to modes with even symmetry about the stellar equator and odd values to modes with odd symmetry. For these modes $n$ gives the number of nodes in $\mathcal{X}_n(\mu)$ over the interval $-1<\mu<1$, except when $-1<x^{-1}<0$ (where $x^{-1}=\bar{\sigma}/2\Omega_s$) when there appears an extra set of nodes. This is related to the existence of quasi g-modes with $n=-1$, see below.

For $n<0$ and $x<0$ the eigenvalues are positive for $|x^{-1}| \rightarrow 0$, and negative for large values of $|x^{-1}|$. In this case the eigenvalue $\lambda_{n m}$ changes sign for $x$-value:
\begin{equation} x^{-1}_r=\frac{- m}{(m+|n|)\,(m+|n+1|)}  \label{x_r} \end{equation}
where  $m>0$.
The oscillation modes corresponding to these roots $x_r$ are the purely toroidal r-modes  (Papaloizou \& Pringle \cite{pap78}). The $n<0$ branches thus have positive, although small, eigenvalues $\lambda_{n m}$  for $|x^{-1}|<|x^{-1}_r|$ and these correspond to quasi-toroidal q-modes (Savonije \& Witte, 2002). Fig.~\ref{lambda} shows the ($\lambda>0$) q-mode branches with $n=-1$ and $n=-2$ for $m$=2 together with the normal g-mode $n$=0 branch. 
The special case $n=-1$ is interesting as this corresponds to a branch of q-modes where $\lambda$ does not remain small when $|x^{-1}|$ decreases, see Fig.~\ref{lambda}.  For $|x^{-1}|<|x^{-1}_r|$ these $n=-1$ modes have values of $\lambda_{n m}$ that remain an order of magnitude smaller than those of the lowest order ($n=0$) g-modes.  This means that for these special 'quasi g-modes' the confinement to the equatorial regions of the rotating star at lower frequencies $\bar{\sigma}$ is almost absent. We will see that for sufficient rotation speeds  $n=-1$ quasi g-mode oscillations can be destabilized by the opacity bump due to the heavy element ionisation zone, like normal g-modes.

The $n<0$ modes with $\lambda_{n m}<0$ (for both positive and negative $x$-values) correspond to rotationally stabilized g$^{-}$-modes or oscillatory convection (and exists only) in the regions where the  square of the Brunt-V\"{a}is\"{a}l\"{a} frequency $\mathcal{N}^2= g\,\mathcal{A}/\Gamma_1 \leq 0$. In this paper only  modes with $\lambda_{n m}>0$ are explored.

\subsection{Solution of radial oscillation equations}
For a given stellar rotation speed $\Omega_s$ and stellar forcing frequency $\bar{\sigma}$ (in fact for a given $x^{-1}=\bar{\sigma}/2\Omega_s$) we can for each possible combination of $(n,m)$ determine the eigenvalue $\lambda_{n}$ and corresponding eigenfunction $\mathcal{X}_n(\vartheta)$ (and auxilliary function $\mathcal{Q}_n$ from which $\epsilon_n$ is calculated), substitute $\lambda_n$ and $\epsilon_n$ in the radial oscillation equations (\ref{eq2}-\ref{eq3})  and solve the set of radial equations (\ref{eq1}-\ref{eq4}). The solution of these differential equations is found by transforming them into algebraic equations by means of finite differences on a staggered spatial mesh and applying matrix inversion similar to standard Henyey schemes for stellar evolution (e.g. Savonije \& Papaloizou 1983). Hereby the forcing potential $\Phi_T(l,m)$ is chosen to have the correct symmetry, i.e. for given $(n,m)$, we adopt $l$ accordingly: for g-modes $l=m+n$ and for q-modes $l=m+|-n-2|$. By scanning through a range of (initially) real forcing frequencies $\bar{\sigma}$ and solving the radial oscillation equations we find the possible oscillation modes (with a different number of radial nodes $k$) when the response becomes resonant for certain values $\bar{\sigma}_k(\Omega_s,n,m)$.  At resonance (that is what we are interested in) the stellar response becomes that of the free oscillation mode $(k,n,m)$ and appears virtually independent (apart from a fixed overall scaling factor) of the value adopted for $l$ in the forcing potential, as long as the equatorial symmetry of the forcing is consistent.

\subsection{Search for unstable oscillation modes}
The resonances with free oscillation modes are found by varying the real valued $\bar{\sigma}$ and maximizing the absolute value of the tidal torque integral (interpreting the forcing potential as that of a spherical harmonic component of an external point mass companion as in  Savonije \& Witte 2002):
\begin{equation}
  \mathcal{T}_{n m}(\bar{\sigma},\Omega_\mathrm{s})= \pi \, \zeta_{n m} \, f \int_0^{R_\mathrm{s}} 
  \mathcal{I}m \left[\rho'_{n m}(r)\right] r^{l+2} \mathrm{d}r \label{torq} 
\end{equation}
where $\pi$ results after the integration
over $\varphi$,   $f$ is the constant in the tidal potential defined by Eq.~(\ref{eqpot}), 
 $\mathrm{Im}$ stands for imaginary part, and

\[
 \zeta_{n m}=\frac{ \left[ \int_{-1}^1 P_l^m(\mu) \mathcal{X}_n(\mu) \,\mathrm{d}\mu \right]^2} {\int_{-1}^1  \mathcal{X}^2_n(\mu)\, \mathrm{d}\mu} \]
(The tidal torque follows by multiplying $ \mathcal{T}_{n m}$ with $m$).
This procedure shows in fact immediately whether the resonance corresponds to a stable oscillation mode (if $ \mathcal{T}_{n m}$ has same sign as $\bar{\sigma}$) or to an unstable one (when $ \mathcal{T}_{n m}$ has opposite sign as $\bar{\sigma}$). This can be understood physically by considering a stable oscillation mode: when the forced oscillation is prograde ( $\bar{\sigma}>0$) the torque should lead to a spin up of the star and must be positive. For an unstable mode the phase difference between forcing and star has the opposite sign. 
 
For a check of the stability properties of a mode and in order to estimate its growth or damping rate, we forced the star with a complex frequency $\sigma$ and searched with a numerical algorithm for that value of the imaginary part $\mathcal{I}m(\sigma)$ for which the integral over the star $\mathcal{J}=\int_0^R \xi_r(r) \xi^*_r(r) \, \mathrm{d} r$ becomes maximized. The search interval for  $\mathcal{I}m(\sigma)$ was chosen symmetric about the zero value.  The resonant forcing of stable oscillation modes becomes enhanced when the radiative damping is neutralized, so that $\mathcal{J}$ becomes maximal for a positive value of $\mathcal{I}m(\sigma)$ (identified as the `damping rate'), in analogy with the standard theory for a 1D forced linear harmonic oscillator with damping. For an unstable mode with `negative damping', the enhanced resonant response is found for $\mathcal{I}m(\sigma)<0$ (the absolute value of which is identified as the `growth rate'). Except for modes near the boundaries of the instability bands,  the integral $\mathcal{J}$ of the squared oscillation amplitude becomes many orders of magnitude larger when applying  complex forcing frequencies to search for the maximum resonant response. 

Please note that in this paper we consider  $\sigma$ to be the {\it real part} of the forcing frequency,  unless its complex character is made explicit, like in  $\mathcal{I}m(\sigma)$,  while $\bar{\sigma}$ always refers to the real (part of the) oscillation frequency (in the corotating frame).

\subsection{$\kappa$-mechanism}
The instabilities found in this paper are (checked to be) caused by the well known $\kappa$ valve-mechanism by which a fraction of the outflowing stellar thermal energy-flux is tapped by an increase of the radiative opacity during the {\it compression} phase of the oscillations and heat is released during the expansion phase. According to a semi-adiabatic analysis assuming constant radiative luminosity the opacity $\kappa$ has the correct absorbing behaviour in regions of the star where (e.g. Unno et al. \cite{unno89}):
\begin{equation} 
\frac{\mathrm{d}}{\mathrm{d r}} \left(\kappa_T + \frac{\kappa_\rho}{\Gamma_3 -1} \right) > 0 
\label{kappa_inst}
\end{equation}
Regions where inequality (\ref{kappa_inst}) is fulfilled tend to drive the oscillation when a sufficient amount of thermal energy can be absorbed during the compression phase. For this the oscillation period should be comparable to the thermal timescale in the driving zone. Let us define a local {\it 'adiabatic frequency'} $\nu_{ad}=2 \pi/\tau_{leak}$ and use the leaking time $\tau_{leak} \simeq  (R_s-r)^2 \,\rho \kappa/c$  as a crude estimate for the thermal timescale $\tau_{th}$ at radius $r$ based on random walk of photons towards the stellar surface. When the surface is approached the leaking time becomes ever shorter and $\nu_{ad}$ rises steeply, see e.g. Fig.~\ref{v04}. This figure shows the various frequencies as a function of the temperature in a 4 $M_\odot$ main-sequence star with Z=0.02.

Note that from now on we normalize all frequencies on the star's critical rotation frequency $\Omega_c=\sqrt{G\, M_s/R_s^3}$ also when this is not indicated. 
In the surface regions where the oscillation frequency becomes much smaller than the `adiabatic frequency' $\nu_{ad}$  the oscillations evidently are very non-adiabatic and no driving can occur when (\ref{kappa_inst}) attains positive values in these regions. For instability the rise of  $\nu_{ad}$, where its curve in Fig.~\ref{v04} intersects the actual (constant) $\bar{\sigma}$ should occur well into the driving region where (\ref{kappa_inst}) attains a significant positive value, i.e. in the opacity bump region near $T\simeq 2 \times 10^5$K, or rise steeply early in the damping region adjacent to it (so that the extra damping has no effect due to the already strong non-adiabaticity).

It is known (Berthomieu et al. \cite{berth78}, Savonije et al. \cite{sav95}) that the traditional approximation yields qualitatively correct results for g-modes if the oscillation frequency $\bar{\sigma}$ is small compared to the Brunt-V\"{a}is\"{a}l\"{a} frequency (thus outside the convective core), so that the horizontal oscillatory motion dominates. This is also a reasonable approximation for the q-modes with weak radial motion studied here. Fig.~\ref{v04} shows that this requirement is amply satisfied for the 4$M_\odot$ star.

\subsection{Comparison with other stability analyses}
We applied our calculation method to a non-rotating star and compared the results with those given in Gautschy and Saio (\cite{gautschy96}) and found an instability region for a 5$M_\odot$ MS star very similar to the one shown in their Figure 2. 
We also compared our results for {\it rotating} stars with those of  Lee (\cite{lee01}) who used a truncated series expansion in spherical harmonics to describe the non-adiabatic non-radial oscillations in a rotating star. Lee  applied this scheme to a $4M_\odot$ ZAMS star with $X_c$=0.7 and $Z$=0.02. We will compare the results for rotation speeds $\Omega_s$=0.1 (Table~\ref{compare_1}) and  $\Omega_s=0.3$ (Table~\ref{compare_2}). Our 4$M_\odot$ ZAMS stellar input model has $^{10} \log(L/L_\odot)$=2.366 and  $^{10} \log T_\mathrm{eff}$=4.170 and thus is slightly hotter than Lee's input model ($^{10} \log T_\mathrm{eff}$=4.164).
\begin{table}[htbp]
  \caption{Comparison with Lee's (2001) result for the radial orders  $(k_{min},k_{max})$ of unstable g-modes for a $4M_\odot$ ZAMS star ($Z$=0.02), rotating at angular velocity $\Omega_s$=0.1. The upper pair  $(k_{min},k_{max})$ corresponds to the results obtained here, the lower pair are those derived by Lee. For easy comparison we have given the $(l,m)$ values with 
$l=n+|m|$.}
\[
   \begin{array}{|c|c|c|c|c|c|c|}  \hline 
  \mbox{degree} & m=-1 & m=-2 & m=-3 &  m=1 & m=2 & m=3 \\
\hline
\hline
l=1 & (10,14) & .. &  .. & (9,23)  & .. & ..\\
    & (9,15) & .. & ..  & (9,23)  & .. & ..\\
\hline
l=2 & (9,19) & (9,19) & .. & (9,23) & (9,22) & ..\\
    & (6,20) & (8,20) & .. & (9,22) & (9,22) & ..\\
\hline
l=3 & (10,24) & (10,23) & (10,23) & (10,24)& (10,24) & (10,24)\\
    & (9,24) & (9,23) & (9,23) & (9,24) & (9,24) & (9,24) \\
\hline
  \end{array} 
\] \label{compare_1}
\end{table}
\begin{table}[htbp]
  \caption{Comparison with Lee's (2001) result for the radial orders of unstable g-modes for a $4 M_\odot$ ZAMS star ($Z$=0.02), rotating at angular velocity $\Omega_s$=0.3.}
\[
   \begin{array}{|c|c|c|c|c|c|c|}  \hline 
  \mbox{degree} & m=-1 & m=-2 & m=-3 &  m=1 & m=2 & m=3 \\
\hline
\hline
l=1 & (10,13) & .. &  .. & (9,26)  & .. & ..\\
    & (4,15) & .. & ..  & (11,24)  & .. & ..\\
\hline
l=2 & (9,22) & (9,19) & .. & (9,26) & (9,26) & ..\\
    & (7,23) & (6,20) & .. & (7,22) & (8,22) & ..\\
\hline
l=3 & (10,26) & (10,25) & (10,22) & (10,27) & (10,27) & (10,26)\\
    & (7,21) & (9,24) & (8,23) & (8,20) & (9,20) & (10,24) \\
\hline
  \end{array} 
\] \label{compare_2}
\end{table}
The results presented in Tables \ref{compare_1}-\ref{compare_2} show that the two methods yield very similar instability intervals for a 4$M_\odot$ ZAMS star. Only for the highest rotation rate we find slighly narrower instability domains for prograde g-modes on one hand and slighly more extended (to lower frequencies) intervals for retrograde g-modes on the other. This may be partly due to the slightly different input model.

It makes practically no difference to these results whether we adopt $\epsilon_n$=0 or $\epsilon_n \neq 0$, in the former case the instability interval is, for the case $\Omega_s$=0.3, sometimes extended whereby $k_{min}$ decreases by one. The great benefit of the present calculation method is that there is no truncation problem in the solution of the angular eigenfunctions  (although by neglecting the horizontal component of $\Omega_s$, which seems reasonable for the considered modes). The horizontal part of the energy flux perturbation is treated only approximately, see equation (\ref{D_a}). But, since the approach in this paper greatly simplifies the analysis of the stability of oscillation modes in rotating stars and provides us with an at least qualitatively correct picture of the linear stability properties of these modes, it seems a good initial approach. Townsend (\cite{town05}) used a similar approach (but neglected the horizontal diffusion altogether by  adopting $\epsilon_n=0$) to study the stability of g-modes in rotating SPB-stars in more detail than is done here. Where the results overlap they seem qualitatively consistent with each other, although Townsend finds unstable $n$=0 g-modes even in ZAMS stars somewhat less massive than 3$M_\odot$ while the $n$=0 g-modes appear all stable in our 3$M_\odot$ models, see below. This seems related to the higher effective temperatures of the lower mass stellar models used by Townsend as compared to the models used here.

\section{Results}

\subsection{Unstable $n=-1$ q-modes in a $3 M_\odot$ MS star (Z=0.02) \label{lab_030}}
\begin{figure}[htbp]
  \includegraphics{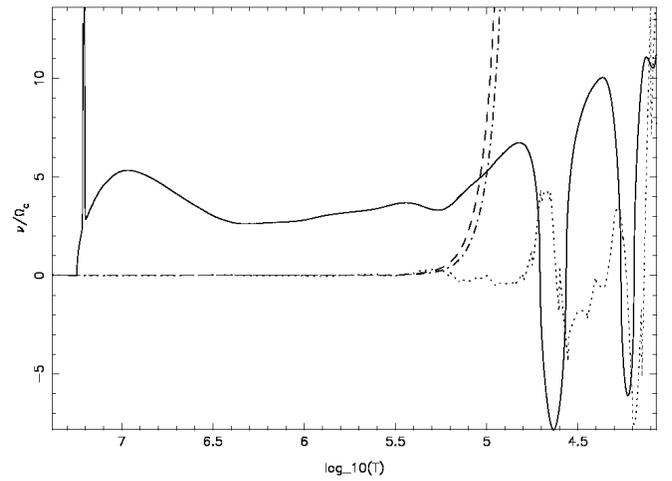}
 \caption{MS star of 3 $M_\odot$ with $X_c$=0.6 and Z=0.02. {\it Full line}: the modified Brunt-V\"{a}is\"{a}l\"{a} frequency  $\nu_\mathrm{BV}$=$\,\mathrm{sign}(\mathcal{N}^2) \sqrt{|\mathcal{N}^2|}$, where $\mathcal{N}$ is the Brunt-V\"{a}is\"{a}l\"{a} frequency; {\it dashed line}: the `adiabatic frequency' $\nu_{ad}=2\, \pi/\tau_{leak}$; {\it dot-dashed line}: again $\nu_{ad}$, but now for $X_c=0.4$;  {\it dotted line}: the value of the radial derivative in expression (\ref{kappa_inst})  in arbitrary units,  all versus temperature.} 
\label{v03} 
\end{figure}

Before exploring the stability of more massive MS stars let us search for the lower mass limit for which overstable q-modes can exist. We studied the stability of two MS models with masses $3M_\odot$ and $2M_\odot$. For the  $2 M_\odot$  MS star no q-mode instabilities (nor any unstable $n$=0 g-modes) could be found, so that the lower boundary of the q-mode instability region lies between 2$M_\odot$ and  3$M_\odot$, coinciding by the way with the lower boundary for SPB stars. It can be seen in Fig.~\ref{v03} that for these lower stellar masses the Z-bump in the opacity near $^{10}\log T \simeq 5.3$ is rather modest and because it lies quite deep inside the star, where the local thermal timescale $\tau_{th}$ is relatively long, only very low frequency oscillation modes can be sufficiently non-adiabatic in this potential driving zone to give rise to overstability.  It appears that the very low frequency quasi g-modes (or $n$=-1 `q-modes') can indeed be excited in the 3$M_\odot$ star, see Table~\ref{tq-1_03} for a list of unstable modes. No unstable g-modes could be detected in a 3$M_\odot$ MS star: the required low frequencies can only be reached for high radial order modes for which internal damping connected with short wavelengths is too strong.

\begin{figure}[htbp]
  \includegraphics{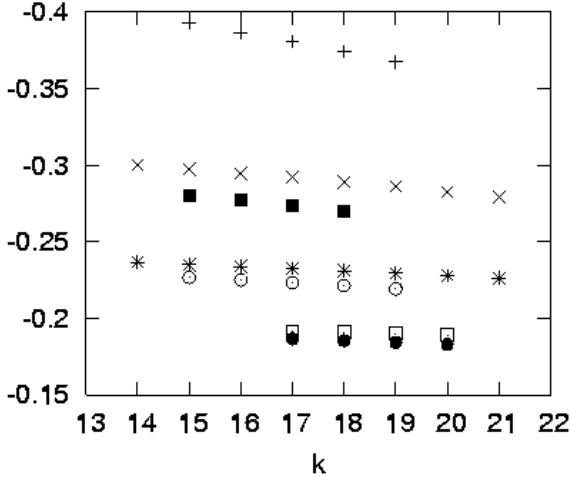}
  \caption{Unstable $n$=-1 q-modes: $\bar{\sigma}_k/(2 \Omega_s)$ versus the number of radial nodes $k$. $\Omega_s$=0.3: {\it plusses}: $m$=1, {\it crosses}: $m$=2, {\it asterisks}: $m$=3 and {\it dotted squares}: $m$=4. Then for $\Omega_s$=0.4: {\it filled squares}: $m$=2;  {\it dotted circles}: $m$=3 and {\it filled circles}: $m$=4. Note that for given $\Omega_s$ the $n$=-1 q-mode frequencies $|\bar{\sigma}_k|$ {\it decrease} with increasing $m$. Model: 3 $M_\odot$, $X_c=0.6$, Z=0.02. } 
  \label{q-1_3060}
\end{figure} 
Fig.~\ref{q-1_3060} shows the oscillation frequencies in the corotating frame: $\bar{\sigma}_k/(2 \Omega_s)$ of $n$=-1 q-modes for several rotation rates of the $X_c=0.6$ and $X_c$=0.4 model. The (absolute value) of the frequencies in the corotating frame $|\bar{\sigma}_k|$  are for $m>1$ significantly lower than those of unstable normal g-modes.  For given $\Omega_s$ the mode frequencies $|\bar{\sigma}_k|$ are  seen to decrease with increasing $m$-value, while for given $m$ the frequencies $|\bar{\sigma}_k|$  increase with rotation speed $\Omega_s$. That $|\bar{\sigma}_k|$ increases with $m$ follows from equation (\ref{x_r}) which tells us that the roots $|x_r^{-1}|$ from where $\lambda_{n m}>0$  for $n=-1$ modes shifts towards 0 (respectively $x^{-1}_r=-1/2$,$ -1/3$ and $-1/4$ for $m$=1, 2 and 3), so that the lowest radial order mode frequency $|\bar{\sigma}_0|$ for a given $m$ is larger  than that for $m+1$ and consequently $|\bar{\sigma}_k(m)|> |\bar{\sigma}_k(m+1)|$ for not too large $k$.

For $m>1$ the mode frequencies of adjacent radial orders $k$ are very densely packed, so that the oscillation periods $P_k$ of {\it all} q-modes in an instability interval $(k_{min},k_{max})$ can only be distinguished if the observations have good time coverage, see Table~\ref{tq-1_03}. This property of q-modes tends to give a much simpler observable oscillation spectrum in rotating stars than g-modes. Note also that although in the corotating frame the q-modes are retrograde modes, in the observer's frame they are always {\it prograde}.

The location and extent of the instability intervals are very sensitive to the thermal timescale $\tau_{th}$ in the opacity Z-bump region. In the more evolved MS models, with cooler surface layers, this region lies deeper inside the star so that $\tau_{th}$ is longer and the steep rise of the 'adiabatic frequency' curve takes place a bit further outwards in the star, at lower $T$, see dot-dashed curve in Fig.~\ref{v03}. Instabilities in the less evolved $X_c$=0.6 model with the shorter $\tau_{th}$ require somewhat higher oscillation frequencies for instability and can thus occur at higher rotation rates than in the more evolved $X_c$=0.4 model, see Table~\ref{tq-1_03}.  For rotation speed $\Omega_s$=0.3 instabilities can only be found  in the $X_c$=0.4 model for $m\geq 3$ (i.e. for the lowest mode frequencies), while for still higher rotation rates (and thus higher mode frequencies) no q-mode instabilities could be found.
We see in Table~\ref{tq-1_03} that the modes with higher $m$ values (i.e. with lower frequencies $|\bar{\sigma}_k|$) become stable when $\Omega_s$ decreases as the oscillations become too slow and hence too non-adiabatic.
\begin{table}[htbp]
\caption{Unstable $n=-1$ q-modes for a 3$M_\odot$ star with Z=~0.02 at two evolution stages indicated by the core hydrogen mass fraction (between curly brackets). The unstable radial orders  $(k_{min},k_{max})$ for different rotation rates $\Omega_s$(in units of $\Omega_c$)  and $m$-values are listed between round brackets . The numbers between square brackets list the oscillation periods (days) corresponding to $k_{min}$ and $k_{max}$ in the observer's frame. Positive periods correspond to prograde (in observer's frame) modes, negative periods to retrograde modes. Empty slots mean that no unstable modes could be detected. $P_s$ is the stellar rotation period. For $X_c=0.4$ and $\Omega_s=0.4$ no unstable modes were detected.}
\[
\begin{array}{|c|c|c|c|c|c|}  
\hline
 \Omega_s &  P_s(d) & m=1 & m=2 & m=3 & m=4 \\
\hline
\hline 
0.4 & 0.57 &  & (15,18) & (15,19) & (17,20) \\
    & \{0.6\} & & [.396,.390] & [.224,.223] & [.157,.157] \\
\hline
0.3 & 0.76 & (15,19) & (14,21) & (14,21) & (17,20)\\
    & \{0.6\} & [3.6,2.9] & [.54,.53] & [.301,.298] & [.210,.210]  \\
\hline
0.2 & 1.14 & (13,22) & (14,22) & (15,20) &  \\
    & \{0.6\} & [11.,6.] & [0.83,0.81] & [.453,.452] & \\
\hline
0.1 & 2.28 & (13,21) & & & \\
    & \{0.6\} & [79.,36.] & & & \\
\hline
\hline
 0.3 & 1.11 &  &  & (25,27) & (25,26) \\
 & \{0.4\} &  &  &  [0.438.,0.437] & [.307,.307] \\
\hline
0.2 & 1.66 & (21,27) & (21,31) & (21,30) \\
 & \{0.4\} & [15.9,10.9] & [1.22,1.20] &  [.662,.660]\\
\hline
0.1 & 3.33 & (20,32) & (22,29) & & \\
&  \{0.4\} & [121.,53.] & [2.48,2.47] & & \\
\hline
  \end{array} 
\] \label{tq-1_03}
\end{table}

\subsection{Unstable modes in a $4M_\odot$ evolved MS-star}
\subsubsection{Unstable retrograde $n$=0 g-modes} 
\begin{figure}[htbp]
  \includegraphics{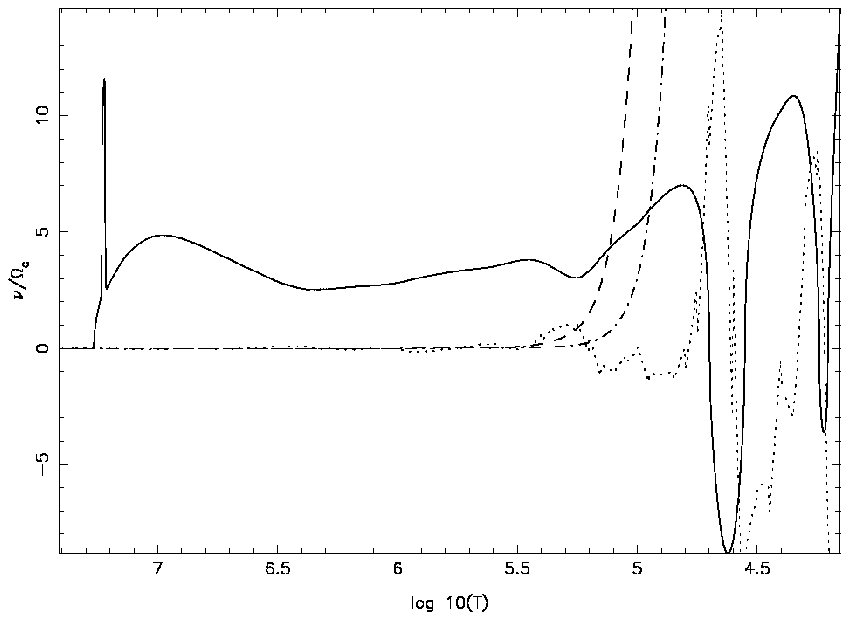}
  \caption{MS star of 4$M_\odot$ with $X_c$=0.6 and Z=0.02. {\it Full line}: the modified Brunt-V\"{a}is\"{a}l\"{a} frequency  $\nu_\mathrm{BV}$=$\,\mathrm{sign}(\mathcal{N}^2) \sqrt{|\mathcal{N}^2|}$, where $\mathcal{N}$ is the Brunt-V\"{a}is\"{a}l\"{a} frequency; {\it dashed line}: the `adiabatic frequency' $\nu_{ad}=2\, \pi/\tau_{leak}$; {\it dot-dashed line}: again $\nu_{ad}$, but now for $X_c=0.2$;  {\it dotted line}: the value of the radial derivative in expression (\ref{kappa_inst})  in arbitrary units,  all versus temperature.} 
  \label{v04} 
\end{figure}

\begin{figure}[htbp]
  \includegraphics{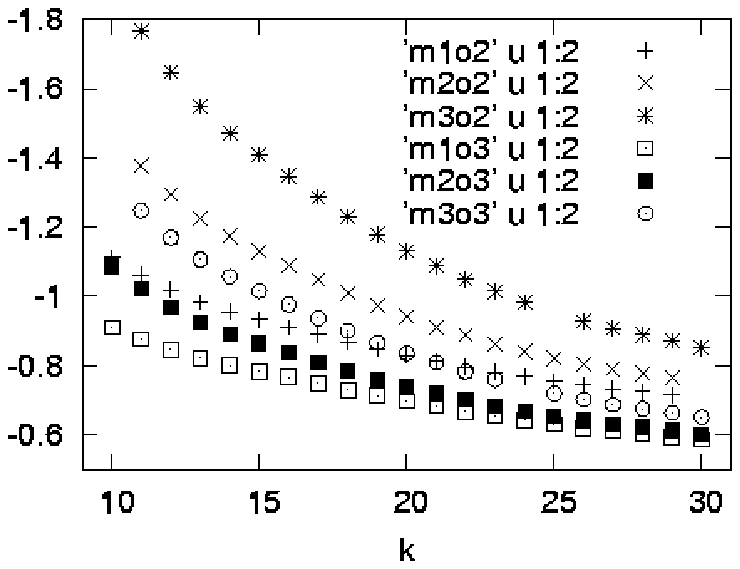}
  \caption{Unstable $n=0$ retrograde g-modes: $\bar{\sigma}_k/(2 \Omega_s)$ versus radial node number $k$. For meaning of symbols see legend above, whereby 'm1o2' denotes case with $m$=1 and $\Omega_s$=0.2, etc.  For given $\Omega_s$ the absolute value of the mode frequencies $|\bar{\sigma}_k|$ increase for increasing $m(=l)$. Model: 4$M_\odot$ with $X_c$=0.6 and Z=0.02.}
  \label{g4}
\end{figure}

Let us now turn to 4 $M_\odot$ MS star models of decreasing central hydrogen abundance $X_c$.
Fig.~\ref{v04} shows the (modified) Brunt-V\"{a}is\"{a}l\"{a} frequency, the `adiabatic frequency' and the driving/damping zones, all as a function of temperature, for the slightly evolved 4$M_\odot$ MS model with $X_c=0.60$. The opacity Z-bump near $^{10} \log{T} \simeq 5.3$ coincides with the region where non-adiabicity of low-frequency, $|\bar{\sigma}/(2 \Omega_s)|< 1$, oscillations becomes noticeable. A little further out the driving is, according to inequality (\ref{kappa_inst}), replaced by damping but the extra thermal energy exchange for relevant frequencies is small because of existing strong non-adiabaticity. By comparing the dashed and dot-dashed curves it is seen that the thermal timescale $\tau_{th}$ in the driving zone increases significantly when the star evolves further from the ZAMS and its surface layers become cooler.
We determined the unstable n=0 g-modes with $m$=1 to $m$=4 in a 4$M_\odot$ main-sequence star rotating at various angular speeds, for two evolution phases: $X_c$=0.6 and $X_c$=0.2, see Table~\ref{tg+0_04}. 
Note that the (retrograde) g-mode frequency $|\bar{\sigma}_k|$ increases with $\Omega_s$ as the increasing Coriolis force renders the star `stiffer' against oscillatory motions.
In Fig.~\ref{g4} it can be seen that, for a given rotation rate $\Omega_s$, the $n$=0 g-mode frequencies $|\bar{\sigma}_k|$ {\it increase} with increasing $m$-values, contrary to the behaviour of $n=-1$ q-modes in Fig.~\ref{q-1_3060}. This is the usual g-mode behaviour, remember that $n$=0 means $m$=$l$ in the non-rotating star terminology. The results in Table~\ref{tg+0_04} show that in the little evolved MS model all combinations of $\Omega_s$ and $m$ yield unstable g-mode intervals. However, for the evolved model with $X_c=0.2$ the opacity bump region has moved into the star with a corresponding longer local value for $\tau_{th}$. As a result the parameter combinations that correspond with higher mode frequencies, i.e. the larger $m$ values and higher rotation rates, no longer yield unstable g-modes because the oscillations have become too adiabatic for destabilization by the $\kappa$-mechanism. For $\Omega_s$=0.5 all considered g-modes appear stable.
\begin{table}[htbp]
  \caption{Unstable $n$=0 retrograde g-modes for a 4 $M_\odot$ star with Z=~0.02.
See Table~\ref{tq-1_03} for explanation.}
\[
   \begin{array}{|c|c|c|c|c|c|}  \hline 
 \Omega_s &  P_s(d) & m=1 & m=2 & m=3 & m=4 \\
\hline
\hline 
0.4 & 0.63 & (10,31) & (10,31) & (11,31) & (12,30) \\
 & \{0.6\} & [-1., -158.] & [3.1,.64] & [.62,.33] & [.35,.22] \\
\hline
0.3 & 0.84 & (10,30) & (10,30) & (11,30) & (12,30) \\
 & \{0.6\} & [-1.,-5.] & [-4.9,1.1] & [1.7,.50] & [.70,.33] \\
\hline
0.2 & 1.27 & (10,29) & (11,29) & (11,30) & (12,30) \\
 & \{0.6\} & [-1.,-2.9] & [-1.7,27.] & [-2.4,.97] & [-29.,.62] \\
\hline
0.1 & 2.53 & (10,24) & (10,27) & (11,29) & (12,30) \\
 & \{0.6\} & [-.99,-2.3] & [-.74,-5.1] & [-.68,-202.] & [-.60,6.1] \\
\hline
\hline
0.4 & 1.45 & (37,44) &  &  &\\
 & \{0.2\} & [-3.3,-4.3] & & &\\
\hline
0.3 & 1.93 & (36,53) & (37,48) & & \\
 & \{0.2\} & [-3.,-4.9] & [11.5,4.5] & \\
\hline
0.1 & 5.79 & (30,67) & (35,65) & (42,49)&  \\
& \{0.2\} & [-2.4,-5.6] & [-2.3,-8.8] & [-2.8,-4.1] & \\
\hline
  \end{array} 
\] \label{tg+0_04}
\end{table}

\subsubsection{Unstable $n=-1$ q-modes}
Let us now consider the behaviour of the $n=-1$ q-modes in a 4$M_\odot$ main-sequence star.
We have extended the calculations to a rather high angular rotation speed of $\Omega_s=0.5$ for which the neglect of centrifugal distortion is questionable just to see the effect of stronger Coriolis forces on the oscillation.
\begin{figure}[htbp]
  \includegraphics{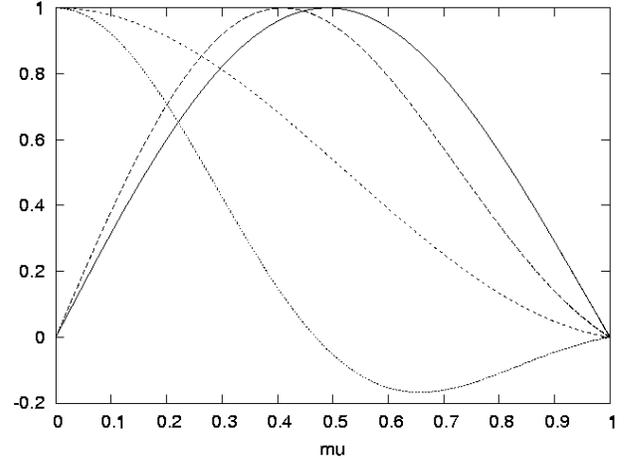}
  \caption{The angular eigenfunction $\mathcal{X}_n(\vartheta)$ versus $\mu=\cos{\vartheta}$ for four different $m$=2 modes, with respectively $n=-1$ and $n$=0, that bracket two instability domains for $\Omega_s=0.3$, see Tables~\ref{tg+0_04}-\ref{4msunq}. From top to bottom: Full line: $q_{12}$; long dashes: $q_{21}$; short dashes: g$_{10}$ and dotted line: g$_{30}$ (the latter shows the extra node and rotational confinement) all as listed in Tables~\ref{tg+0_04} and \ref{4msunq}. The normalisation of y-axis is arbitrarily. Model: 4 $M_\odot$, $X_c=0.6$, Z=0.02.}
  \label{eigenf}
\end{figure} 
Fig.~\ref{eigenf} shows the eigenfunction $\mathcal{X}_n(\mu)$ for the first and last unstable $m$=2 $n$=0 g-mode and $m$=2 $n=-1$ q-mode, bracketing their instability domains, as listed in Tables~\ref{tg+0_04} and \ref{4msunq} for $\Omega_s=0.3$. The two bracketing unstable g-modes are at the boundary, respectively well inside the inertial range: with $1/x\simeq -1.09$ and -0.60, respectively. For the latter low frequency g$_{30}$-mode  rotational effects are thus significant and its eigenfunction $\mathcal{X}_n$ exhibits the typical equatorial confinement of these modes: the eigenfunction has significant amplitude only for $|\mu|< 1/x$.  The two q-modes, by definition inside the inertial regime, have respectively the values $1/x \simeq -0.302$ and $-0.271$, but are seen to exhibit hardly any confinement toward low values of $\mu$ (these q-modes are odd and vanish at the equator).
\begin{table}[htbp]
  \caption{Unstable $n=-1$ q-modes for a 4 $M_\odot$ star with metal content Z=~0.02, see Table~\ref{tq-1_03} for explanation.}
\[
   \begin{array}{|c|c|c|c|c|c|}  \hline 
 \Omega_s &  P_s(d) & m=1 & m=2 & m=3 & m=4 \\
\hline
\hline 
0.5 & 0.51 & (10,27) & (11,25) & (12,22) & (13,21)\\
& \{0.6\}  & [1.8,1.0] & [0.35,0.32] & [.198,.194] & [.139,.138]\\
\hline
0.3 & 0.84 & (11,22) & (14,21) & (14,17) & \\
 & \{0.6\} & [4.9,2.5] & [0.60,0.58] & [.333,.332] & \\
\hline
0.2 & 1.27 & (11,20) & & &\\
 & \{0.6\} & [13.4,6.1]& & & \\
\hline
0.1 &  2.53 &  & & &\\
 & \{0.6\} & & & & \\
\hline
\hline
0.5 & 0.74 & (16,43) & (16,41) & (17,39) & (19,36)\\
& \{0.4\}  & [2.6,1.5] & [0.52,0.47] & [.291,.283] & [.204,.202]\\
\hline
0.3 & 1.23 & (15,39) & (16,35) & (18,25) & (21,25)\\
 & \{0.4\} & [8.6,3.4] & [0.90,0.84] & [.489,.486] & [.3416,.3412]\\
\hline
0.1 & 3.70 & (19,23) & & &\\
 & \{0.4\} & [118,84]\\
\hline
\hline
0.5 & 1.16 & (28,65) & (28,66) & (30,65)  & (31,62)\\
 & \{0.2\} & [4.1,2.4] & [0.81,0.74] & [0.45,0.44] & [.319,.316]\\
\hline
0.3 & 1.93 & (27,66) & (28,63) & (30,58) & (32,53)\\
 & \{0.2\} & [13.1,5.3] & [1.4,1.3] & [0.77,0.75] & [.534,.532]\\
\hline
0.1 & 5.79  & (28,50) &  &  &\\
 & \{0.2\} & (253,88] &  & & \\
\hline
  \end{array} 
\] \label{4msunq}
\end{table}
The q-modes, in spite of their weak compressions also unstable due to the $\kappa$-mechanism, extent the instability region down to lower frequencies than that of the normal g-modes of Table~\ref{tg+0_04}. The q-modes have significantly smaller $\lambda$-values and their  angular eigenfunctions do not exhibit the strong rotational confinement to the stellar equator for small values of $|1/x|$ shown by normal retrograde  g-modes.
\begin{figure}[]
  \includegraphics{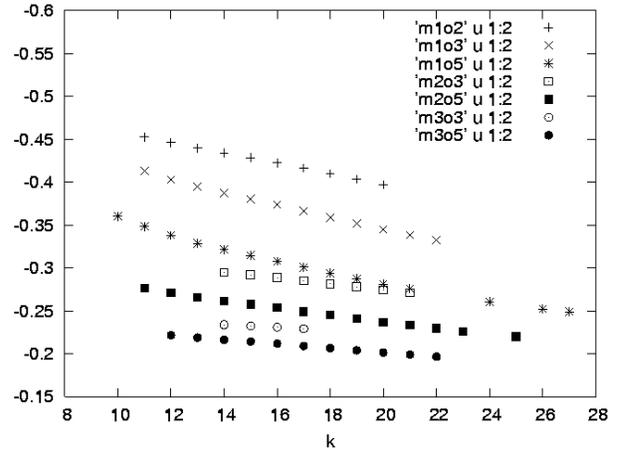}
  \caption{Unstable $n=-1$ q-modes: $\bar{\sigma}_k/(2 \Omega_s)$ versus radial node number $k$. For meaning of symbols see legend above, whereby 'm1o2' denotes case with $m$=1 and $\Omega_s$=0.2, etc. Contrary to the g-modes in Fig.~\ref{g4} the absolute value of the mode frequencies $|\bar{\sigma}_k|$ {\it decrease} for increasing $m$ for a given $\Omega_s$, see text. Model: 4$M_\odot$ with $X_c$=0.6 and Z=0.02.}
  \label{q4}
\end{figure}

In Table \ref{4msunq} we list the unstable q-modes with $n=-1$ and $m$=1 to 4 at various evolutionary stages on the main-sequence and for angular rotation speeds up to $\Omega_s$=~0.5. 
The r-mode frequencies are proportional to $\Omega_s$, and the same remains (roughly) true for the q-modes, although non-toroidal effects begin to  become significant with increasing $k$.
 
In order to understand the shifting boundaries of the instability regions with $m$ and $\Omega_s$ one should again realize that for a given rotation speed $\Omega_s$ the absolute value of the oscillation frequency $|\bar{\sigma}_k|$  for $n=0$ g-modes increases with $m(=l)$, see Fig.~\ref{g4}, while Fig.~\ref{q4} shows that for $n=-1$ q-modes the reverse is true (for an explanation see section \ref{lab_030}). It can be seen in Fig.~\ref{q4}, e.g. the case with $m=1$ and $\Omega_s$=0.5, that sometimes not all radial orders in an instability domain are unstable according to our criteria (the mode amplitude should increase when the search for the resonant forcing frequency is made in the complex plane). In some cases the resonances remain too weak. 
\begin{figure}[htbp]
  \includegraphics{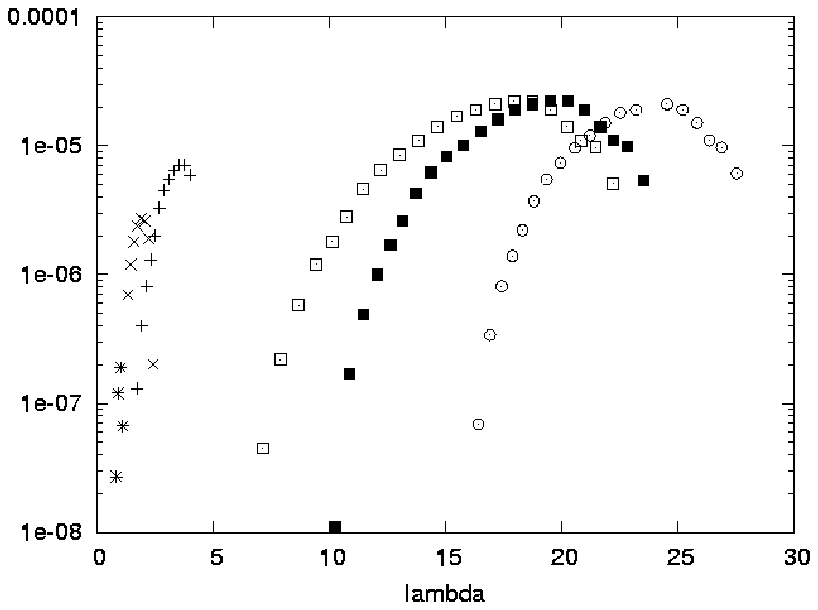}
  \caption{Growth rates $-\mathcal{I}\mathrm{m}(\sigma)$ versus $\lambda$ for unstable $n=-1$ q-modes and $n$=0 g-modes. Symbols:  q-modes with $m$=1,2 and 3: +, x and *; g-modes with m=1,2 and 3: dotted squares, filled squares and dotted circles. Model: 4$M_\odot$ with $X_c$=0.6 and Z=0.02 for $\Omega_s=0.3$.}
  \label{growth_04}
\end{figure}

Fig.~\ref{growth_04} shows the imaginary parts of the complex forcing frequency $\sigma$ for which the resonant response in a 4$M_\odot$ model is maximized for both the unstable $n=-1$ q-modes and $n$=0 g-modes with m-values in the range 1-3. 
 As expected, for the quasi g-modes the highest growth rates are attained by the $m$=1 modes which have the largest $\lambda$ values and highest oscillation frequencies $\bar{\sigma}$ (see Fig.~\ref{q4}) and are thus closest in character to the g-modes. But in this little evolved stellar model ($X_c=0.6$) the driving layer is relatively close to the stellar surface and the (low-frequency) modes  have rather small growth rates. 

In the more evolved stellar model with $X_c$=0.4 and for the lowest rotation speed $\Omega_s$=0.1 the $m$=2 and $m$=3 q-modes have oscillation frequencies too small for the $\kappa$ instability to be effective, the oscillations are too non-adiabatic in the $\kappa$-bump region. In the most evolved stellar model ($X_c=0.1$), on the other hand, the sudden increase of $\tau_{ad}$ occurs further out in the star, see Fig.~\ref{v04}. For $\Omega_s=0.5$ this results in stable $m$=1 q-modes  as the mode frequencies are in this case too high for efficient driving in the opacity bump region.

Note again that for $m\geq 2$ the subsequent radial orders of the $n=-1$ q-modes in an instability interval are densely spaced in period. For normal g-modes similar close spacing of radial orders $k$ generally occurs for much larger values of $k$ for which the oscillations are significantly more confined to the equatorial plane of the rotating star and for which the strong radiative damping prevents instability.

\subsection{Unstable $n=-2$ q-modes in  3-4 $M_\odot$ MS stars}
It appears that in the evolved 3-4$M_\odot$ models even unstable higher order $n=-2$ quasi-toroidal modes are excited by the $\kappa$-mechanism. These modes have very small eigenvalues $\lambda_{-2}$ and very low oscillation frequencies, see Tables~\ref{tq-2_03}-\ref{tq-2_04} and Fig.~\ref{fq-2_03}. G-modes of similarly low oscillation frequencies would be of much higher radial degree and have much larger $\lambda$-values and be stable due to strong non-adiabaticity. In less evolved stellar models with $X_c \geq 0.4$ the $n=-2$ q-modes appear all stable. 
\begin{figure}[htbp]
  \includegraphics{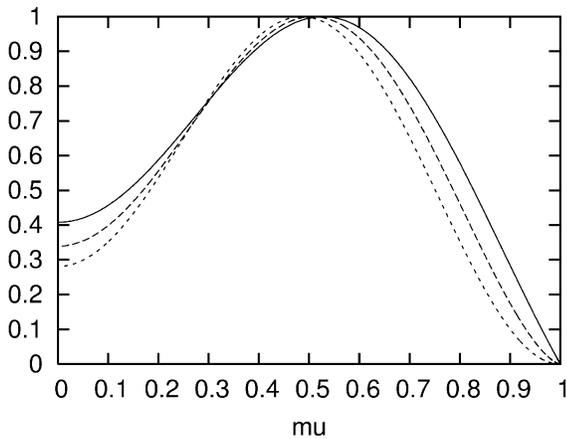}
  \caption{The angular eigenfunction $\mathcal{X}_n(\vartheta)$ versus $\mu$ for the unstable $k$=17 $n=-2$ modes at rotation rate $\Omega_s$=0.4: Solid line: $m$=2; dashed line: $m$=3 and dotted line: $m$=4.  Model: 3 $M_\odot$, $X_c=0.6$, Z=0.02.}
  \label{efq-2_17} \end{figure}
The angular eigenfunction $\mathcal{X}_n(\vartheta)$ for three (unstable)  $n=~-2$ q$_{17}$-modes with $m$=2, 3 and 4, all with frequencies $|1/x|<0.115$ (see  Fig.~\ref{fq-2_03}), are depicted in Fig.~\ref{efq-2_17}. The eigenfunction $\mathcal{X}_n(\vartheta)$ determines the latitudinal distribution of temperature, density (and small radial velocity)  over the stellar surface. The shape varies little as a function of $m$. 
\begin{table}[htbp]
  \caption{Unstable $n=-2$ q-modes for a $3 M_\odot$ MS star with Z=0.02, see Table~\ref{tq-1_03} for explanation. For $\Omega_s$ lower than listed no unstable modes could be found.}
\[
   \begin{array}{|c|c|c|c|c|c|}  \hline
 \Omega_s &  P_s(d) & m=1 & m=2 & m=3 & m=4 \\
\hline
\hline 
0.4 & 0.57 &  & (13,18) & (14,19) & (16,18) \\
 & \{0.6\} &  & [0.32,0.31] & [.206,.204] & [.1509,.1507] \\
\hline
0.3 & 0.76 &  & (14,17) & (16,17) & \\
 & \{0.6\} &  & [.434,.429] & [.2764,.2760] &  \\
\hline
\hline
0.4 & 0.83 & (19,25) & (19,30) & (21,31) & (22,29)\\
 & \{0.4\} & [0.98,0.95]  & [0.47,0.46] & [.301.,.297] & [.221,.220]\\
\hline
0.3 & 1.11 & (20,23) & (20,29) & (21,28) & (24,27)\\
 & \{0.4\} & [1.36,1.32] & [0.64,0.62] &  [.405.,.402] & [.295,.295]\\
\hline
0.2 & 1.66 & (21,22) & (20,26) &   & \\
 & \{0.4\} & [2.17,2.16] & [.975,0.964] &  & \\
\hline
  \end{array} 
\] \label{tq-2_03}
\end{table}
\begin{table}[htbp]
  \caption{Unstable $n=-2$ q-modes for a 4 $M_\odot$ star with $X_c=~0.2$ and metal content Z=~0.02, see Table~\ref{tq-1_03} for explanation. The oscillation periods in the corotating frame (between square brackets) are of order  the rotation period over $m$ due to the small values of $|\bar{\sigma}_k|$.}
\[
   \begin{array}{|c|c|c|c|c|c|}  \hline 
 \Omega_s &  P_s(d) & m=1 & m=2 & m=3 & m=4\\
\hline
\hline 
0.5 & 1.16 &  & (28,43) & (29,47) & (32,47)\\
 & \{0.2\} &  & [0.65,0.63] & [0.418,0.410] & [.306,.304]\\
\hline
0.4 & 1.45 &  & (29,40) & (30,44) & (34,43)\\
 & \{0.2\} &  & [0.82,0.80] & [0.525,0.519] & [.384,.382] \\
\hline
0.3 & 1.93 &  & (32,37) & 37  &\\
 & \{0.2\} &  & [1.106,1.096] & [0.70] &\\
\hline
  \end{array} 
\] \label{tq-2_04}
\end{table}
Figures~\ref{v03} and \ref{v04} show that in the evolved stars the `adiabatic' frequency rises steeply only near the outer edge of the opacity bump region. Low frequency q-mode oscillations apparently can still absorb sufficient heat during compression in the opacity bump zone, at least for rotation speeds $\Omega_s\geq 0.3$. When the opacity derivatives $\kappa_T$ and $\kappa_\rho$ are put to zero the instabilities disappear. 

For the higher stellar masses on the main sequence the opacity bump is closer to the surface and  the non-adiabaticity too strong for destabilization of the very low frequency $n=-2$ q-modes.
\begin{figure}[htbp]
  \includegraphics{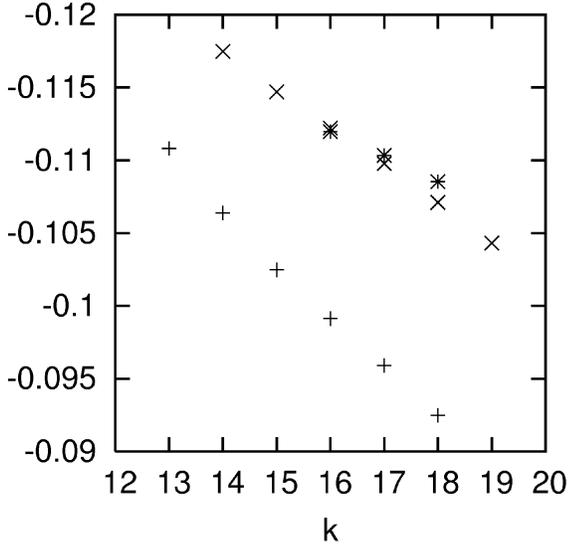}
  \caption{Unstable $n=-2$ q-modes: $\bar{\sigma}_k/(2 \Omega_s)$ versus radial order $k$, for rotation rate $\Omega_s=0.4$: {\it asterisks}: $m$=4, {\it crosses}: $m$=3, {\it plusses}: $m$=2.  Model: 3 $M_\odot$ with $X_c$=0.6 and Z=0.02.}
  \label{fq-2_03}   
\end{figure}
\begin{figure}[htbp]
  \includegraphics{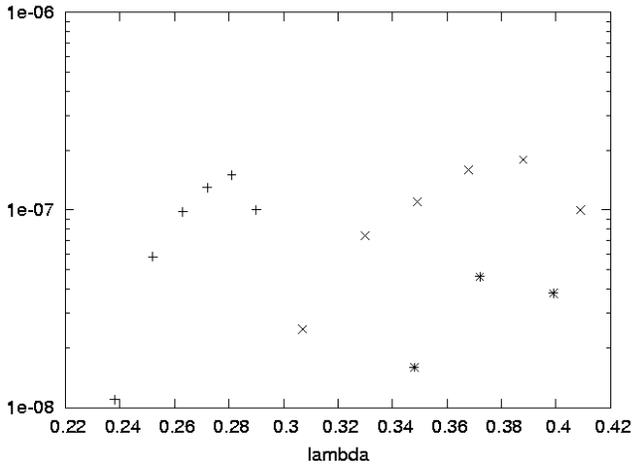}
  \caption{Growth rates $-\mathcal{I}\mathrm{m}(\sigma)$ versus $\lambda$ for the unstable $n=-2$ q-modes shown in Fig.~\ref{fq-2_03}. Symbols:  $m$=2, 3 and 4 are represented by +, x and *, respectively. Model: 3$M_\odot$ with $X_c$=0.6 and Z=0.02.}
  \label{growth_03}
\end{figure}

Fig.~\ref{fq-2_03} shows that the absolute value of the $n=-2$ mode frequencies (in corotating frame) $|\bar{\sigma}_k|$ as a function of azimuthal index $m$ behave differently from the $n=-1$ modes shown in Fig.~\ref{q4}:  the absolute value of the frequencies of e.g. $m$=3 modes are higher than those of $m$=2, while for $k\geq 16$ the same holds for $m$=4 with respect to $m$=3. This can again be explained by considering equation (\ref{x_r}). For $n=-2$ the roots $x^{-1}_r$ have values $-1/6$, $-1/6$, $-3/20$ and $-1/10$ for $m$=1, 2, 3 and 4, respectively, i.e. lie closely together so that already for relatively small $k$ we get the behaviour stated above.

Fig.~\ref{growth_03} shows the imaginary parts of the complex forcing frequency $\sigma$ for which the resonant response is maximized for the unstable $n=-2$ q-modes with m-values in the range 2-4. The real parts of the oscillation frequencies are shown in Fig.~\ref{fq-2_03}. The very low frequency $n=-2$ q-modes with small $\lambda$-values have significantly smaller growth rates than the $n=-1$ q-modes. The $m$=3 modes attain the largest maximum growth rate.

\subsection{Unstable modes in a $6 M_\odot$ MS star}
\subsubsection{Unstable $n$=0 retrograde g-modes for Z=0.02}
\begin{figure}[htbp]
  \includegraphics{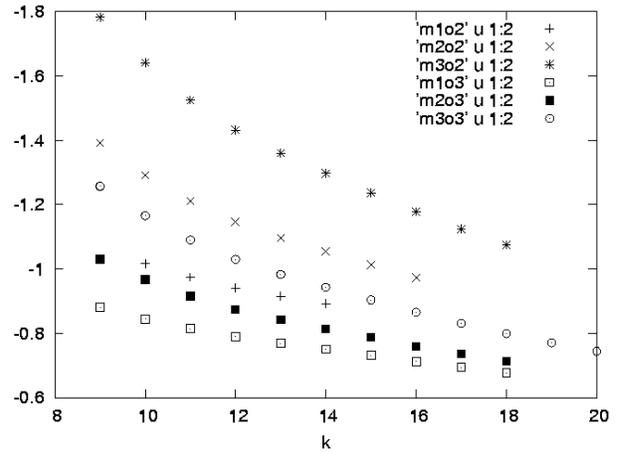}
  \caption{Unstable $n=0$ retrograde g-modes: $\bar{\sigma}_k/(2 \Omega_s)$ versus radial node number $k$. For meaning of symbols, see Figure, whereby 'm1o2' denotes case with $m$=1 and $\Omega_s$=0.2, etc.  For given $\Omega_s$ the absolute value of the mode frequencies $|\bar{\sigma}_k|$ increase for increasing $m$=l. Model: 6$M_\odot$ with $X_c$=0.6 and Z=0.02.}
  \label{fg0_06060}
\end{figure}

Fig.~\ref{fg0_06060} shows the frequencies of unstable $n$=0 g-modes in the 6$M_\odot$ model with $X_c$=0.6, rotating at speeds $\Omega_s$=0.1 and 0.3. By comparing with Fig.~\ref{g4} which depicts the same for a 4$M_\odot$ stellar model it appears that in this more massive star the instability intervals are less extended and terminate at larger $k$-values ($\Delta k \simeq -10$), while the high frequency edge is shifted to slightly lower $k$ with $\Delta k\simeq -2$. This is caused by the shorter thermal timescale in the driving zone of the more massive star. Apart from the slight shift in $k$, the frequencies $\bar{\sigma}_k$ of the unstable g-modes are very similar for the 4$M_\odot$ and 6$M_\odot$ models.
\begin{table}[htbp]
  \caption{Unstable $n$=0 g-modes for a $6 M_\odot$ MS star with Z=0.02, see Table~\ref{tq-1_03} for explanation.}
\[
   \begin{array}{|c|c|c|c|c|}  \hline 
 \Omega_s &  P_s(d) & m=1 & m=2 & m=3 \\
\hline
\hline 
0.3 & 0.98 & (9,18) & (9,18) & (9,20) \\
  & \{0.6\} & [-1.3,-2.8] & [-16.,1.7] & [2.,.65] \\
\hline
0.2 & 1.47 & (10,14) & (9,16) & (9,18) \\
 & \{0.6\} & [-1.4,-1.9] & [-1.9, 28.] & [-2.6,1.7] \\
\hline
 0.1 & 2.53 & (10,24) & (10,27) & (11,29) \\
 & \{0.6\} & [-.99,-2.3] & [-.74,-5.1] & [-.68,-202.] \\
\hline
\hline
0.5 & 0.85 & (12,37) & (12,37) &  (13,38) \\
 & \{0.4\} & [-1.6,11.0] & [2.5,0.8] & [0.7,0.4]\\
\hline
0.3 & 1.55 & (13,31) & (13,32) & (13,34) \\
 & \{0.4\} & [-1.7,-5.0] & [-6.6,2.1] &  [5.5,0.89]\\
\hline
0.1 & 4.26 & (14,20) & (13,25) & (13,29) \\
 & \{0.4\} & [-1.8,-2.5] & [-1.2,-3.5] &  [-0.95,-6.1]\\
\hline
\hline
0.5 & 1.32 & (21,68) & (21,60) & (21,59)\\
 & \{0.2\} & [-2.5,13.2] & [4.0,1.2] & [1.2,0.67]\\
\hline
0.3 & 2.20 & (21,61) &  (21,62) &  (21,63)\\
 & \{0.2\} & [-2.4,-10.4] &  [-6.5,2.9] & [31.0,1.3]\\
\hline
0.1 & 6.59 & (21,42) & (21,50) & (21,57) \\
 & \{0.2\} & [-2.3,-4.7] & [-1.7,-7.6] &  [-1.3,-21.1]\\
\hline
  \end{array} 
\] \label{6msung}
\end{table}
Table~\ref{6msung} lists the g-mode instability intervals for various rotation rates and evolution stages of the 6$M_\odot$ MS star. The table shows that we expect to observe many unstable g-modes with a retrograde character in observer's frame (negative periods). By writing $\sigma=\bar{\sigma} + m\, \Omega_s$ we can understand that the modes become prograde in the observer's frame for larger $m$-values and or higher stellar rotation rates as listed in table~\ref{6msung}.
The unstable g-modes, both prograde and retrograde, occur in many wide instability intervals which is, however, not observed in SPB stars. We list here only the $n$=0 g-modes, but there exist also many higher order ($n>0$) unstable g-modes. It is a serious puzzle why all these apparently unstable g-modes are not observed.
\begin{figure}
  \includegraphics{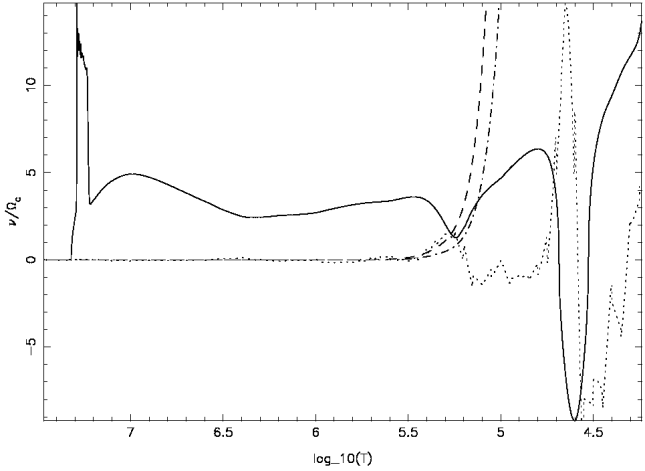}
  \caption{MS star of 6$M_\odot$ with $X_c$=0.6 and Z=0.02. {\it Full line}: the modified Brunt-V\"{a}is\"{a}l\"{a} frequency  $\nu_\mathrm{BV}$=$\,\mathrm{sign}(\mathcal{N}^2) \sqrt{|\mathcal{N}^2|}$, where $\mathcal{N}$ is the Brunt-V\"{a}is\"{a}l\"{a} frequency; {\it dashed line}: the `adiabatic frequency' $\nu_{ad}=2\, \pi/\tau_{leak}$; {\it dot-dashed line}: again $\nu_{ad}$, but now for $X_c=0.4$;  {\it dotted line}: the value of the radial derivative in expression (\ref{kappa_inst})  in arbitrary units,  all versus temperature.} 
  \label{v06} 
\end{figure}

\subsubsection{Unstable $n=-1$ q-modes for Z=0.02}
\begin{figure}[htbp]
  \includegraphics{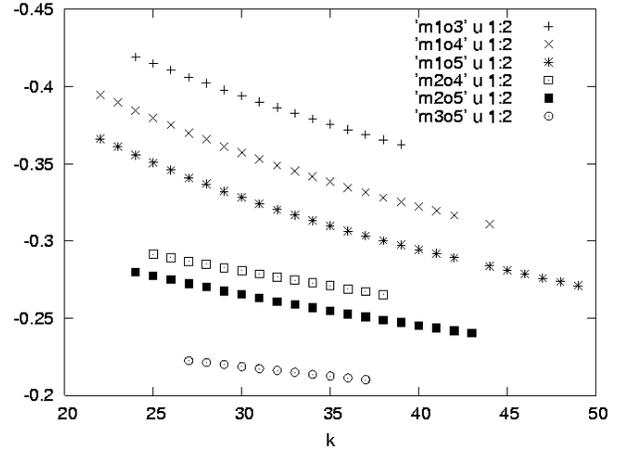}
  \caption{Unstable $n=-1$ q-modes: $\bar{\sigma}_k/(2 \Omega_s)$ versus the number of radial nodes $k$. For meaning of symbols, see Figure, whereby 'm1o3' denotes case with $m$=1 and $\Omega_s$=0.3, etc. Model: 6 $M_\odot$, $X_c=0.2$, Z=0.02. } 
  \label{plot4}
\end{figure}

The frequencies of unstable $n=-1$ q-modes detected for the 6$M_\odot$ model with $X_c$=0.2 are shown in Fig.~\ref{plot4}.  The instability regions ($k_{min}$,$k_{max}$) are smaller than for the g-modes: due to the lower oscillation frequency of q-modes the low-frequency (high k) end is less extended as the opacity bump region becomes too non-adiabatic. Note that all unstable q-modes in Fig.~\ref{plot4} have $|1/x|<0.45$, while many of the unstable g-modes in the same stellar model (except the $m$=1 modes) lie outside the inertial range, see Fig.~\ref{fg0_06060}.
In Table~\ref{6msun1} we list the unstable $n=-1$ modes with $m$=1, 2  or 3 for a $ 6~M_\odot$ main-sequence star. The frequency of the $m$=4 q-modes appears too low for destabilization by the $\kappa$ mechanism. With increasing stellar mass the  opacity bump driving region for the instabilities moves closer to the stellar surface and the steep rise of $\tau_{ad}$  occurs in the 6$M_\odot$ model at higher temperature, see Fig.~\ref{v06}. For the unevolved 6 $M_\odot$ models (with $X_c > 0.4$) the thermal timescale in most of the opacity bump region is therefore shorter than the  oscillation periods (in corotating frame) of the q-modes, except for the fastest rotation rates considered. For this reason we find only $m$=1 instabilities and those only for $\Omega_s\geq 0.4$. When the star evolves away from the ZAMS the effective temperature decreases and the driving zone moves inwards, so that the local thermal timescale increases. For the models with $X_c< 0.4$ we therefore find more unstable q-modes and also instabilities at  lower rotation rates, although not for $\Omega_s<0.3$ as in a 4 $M_\odot$ star: the frequencies are in that case too low.

\begin{table}[htbp]
  \caption{Unstable $n=-1$ q-modes for a $6 M_\odot$ MS star with Z=0.02, see Table~\ref{tq-1_03} for explanation. No unstable $m$=4 modes were found.}
\[
   \begin{array}{|c|c|c|c|c|}  \hline 
 \Omega_s &  P_s(d) & m=1 & m=2 & m=3 \\
\hline
\hline 
0.5 & 0.85 & (15,23) &  &  \\
 & \{0.4\}  & [2.7,2.1] &  & \\
\hline
0.4 & 1.07 & (16,20) &  &  \\
 & \{0.4\} & [4.0,3.3] &  & \\
\hline
\hline
0.5 & 1.32 & (22,49) & (24,43) & (27,37) \\
 & \{0.2\} & [4.9,2.9] & [.92,.87] & [0.516,0.511]\\
\hline
0.4 & 1.65 & (22,44) & (25,38) & \\
 & \{0.2\} & [7.8,4.4] & [1.16,1.12] &  \\
\hline
0.3 & 2.20 & (24,39) & & \\
 & \{0.2\} & [13.6,8.0] & & \\
\hline
  \end{array} 
\] \label{6msun1}
\end{table}

\subsubsection{Unstable $n=-1$ q-modes  for Z=0.03}
When the metal fraction is increased to Z=0.03 the  opacity bump effect gets stronger and more q-modes become unstable, as can be seen by comparing Table~\ref{6msun1} with Table~\ref{6msun2}.
With the enhanced opacity bump the higher $m$ modes can also be destabilized, for high rotation rates $\Omega_s>$0.3 even $m$=4 in the evolved $X_c$=0.2 stellar model.
\begin{table}[htbp]
  \caption{Unstable $n=-1$ q-modes for a $6 M_\odot$ MS star with Z=0.03, see Table~\ref{tq-1_03} for explanation.}
\[
   \begin{array}{|c|c|c|c|c|c|} \hline 
 \Omega_s  &  P_s(d) & m=1 & m=2 & m=3 & m=4\\
\hline
\hline 
0.5 & 0.93 & (14,32) &  (15,27) & (17,24)   &\\
 & \{0.4\} & [3.4,2.0] & [.65,.61]  & [.365,.361] &\\
\hline
0.4 & 1.17 & (14,28) & (16,25) & &\\
 & \{0.4\} & [5.3,2.9] & [.82,.79] & &\\
\hline
0.3 & 1.55 & (15,26) & & &\\
 & \{0.4\} & [9.3,5.3] & & &\\
\hline
\hline
0.5 & 1.47 & (22,56) & (24,58) & (25,53) & (27,47)\\
 & \{0.2\} & [6.0,3.1] & [1.03,0.94] & [.578,.564] & [.406,.403]\\
\hline
0.4 & 1.84 & (22,60) & (24,54) & (26,47) & (32,37)\\
 & \{0.2\} & [9.6,4.2] & [1.31,1.22] & [.726,.714] & [.508,.507]\\
\hline
0.3 & 2.45 & (23,53) & (25,46) & (28,34)  &\\
 & \{0.2\} & [18.,7.2] & [1.77,1.69] & [.971,.968]  \\
\hline
  \end{array} 
\] \label{6msun2}
\end{table}

\begin{figure}[htbp]
  \includegraphics{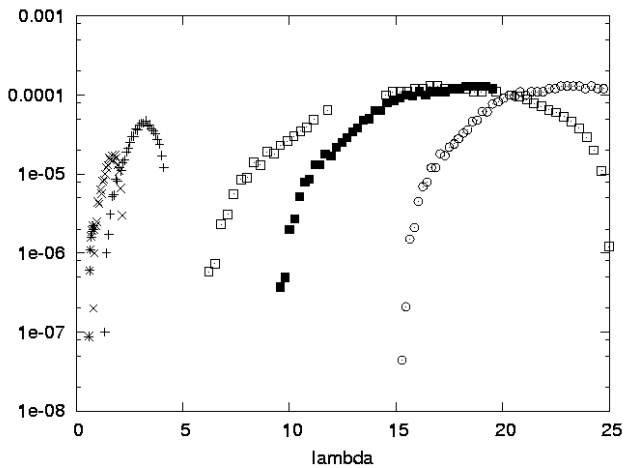}
  \caption{Growth rates $- \mathcal{I}\mathrm{m}(\sigma)$  versus $\lambda$ of unstable $n=-1$ q-modes and $n$=0 g-modes. Symbols:  q-modes with $m$=1,2 and 3: +, x and *; g-modes with m=1,2 and 3: dotted squares, filled squares and dotted circles. Model: 6$M_\odot$ with $X_c$=0.2 and Z=0.03 for $\Omega_s=0.3$.}
  \label{growth_06}
\end{figure}

Fig.~\ref{growth_06} shows the imaginary parts of the complex forcing frequency $\sigma$ for which the resonant response is maximized for both a set of $n=-1$ q-modes and $n$=0 g-modes for various m-values. Again the highest frequency $m$=1 q-modes show the largest growth rates.
It can be seen that the maximum growth rate of $m$=1 q-modes becomes almost comparable ($\sim 20 \%$) to that of $n$=0 g-modes in this more evolved ($X_c=0.2$) main-sequence model.

\subsection{Unstable $n=-1$ modes in a $8 M_\odot$ MS star  with Z=0.03 $\rightarrow$  LPV in Be stars?}
\begin{figure}[htbp]
  \includegraphics{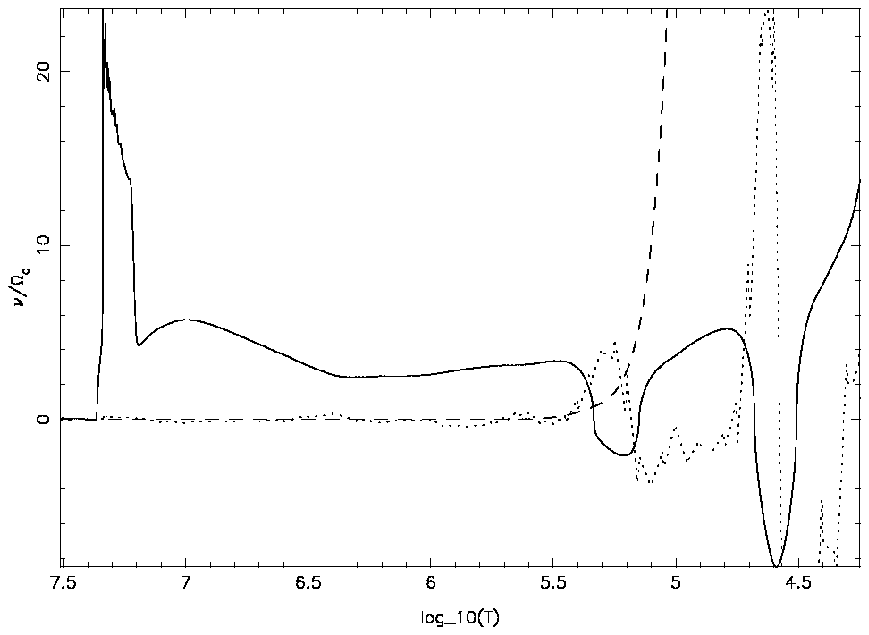}
  \caption{MS star of 8$M_\odot$ with $X_c$=0.2 and Z=0.03. {\it Full line}: the modified Brunt-V\"{a}is\"{a}l\"{a} frequency  $\nu_\mathrm{BV}$=$\,\mathrm{sign}(\mathcal{N}^2) \sqrt{|\mathcal{N}^2|}$, where $\mathcal{N}$ is tfhe Brunt-V\"{a}is\"{a}l\"{a} frequency; {\it dashed line}: the `adiabatic frequency' $\nu_{ad}=2\, \pi/\tau_{leak}$; {\it dotted line}: the value of the radial derivative in expression (\ref{kappa_inst})  in arbitrary units,  versus temperature.}
  \label{V08020} 
\end{figure}

Fig.~\ref{V08020} shows the driving- and damping regions in a 8 $M_\odot$ MS star with $X_c=0.2$ and Z=0.03. It can be seen that for this more massive star the opacity bump region is convectively unstable. Nevertheless, by assuming the convective flux is not heavily perturbed by the oscillation, we have sought for unstable modes, see Table~\ref{8msun}. We have extended the calculations up to completely illegal rotation speeds for which the (neglected) rotational deformation of the star becomes strong in order to compare our results with observations of rapidly rotating Be-stars. 
\begin{table}[htbp]
  \caption{Unstable $n=-1$ q-modes in a 8 $M_\odot$ MS star with $Z=0.03$, see Table~\ref{tq-1_03} for explanation.}
\[
   \begin{array}{|c|c|c|c|c|c|}  \hline 
 \Omega_s &  P_s(d) & m=1 & m=2 & m=3  \\
\hline
\hline 
0.9 & 0.90 & (20,34)    & (21,44) & (23,39) \\
 & \{0.2\} & [2.21,1.70] & [.592,.552] & [.345,.334] \\
\hline
0.7 & 1.15 & (21,37) & (22,38) & (25,33)  \\
 & \{0.2\} & [3.3,2.4] & [.78,.74] & [.446,.442] \\
\hline
0.6 & 1.34 & (21,36) & (23,35)  & \\
 & \{0.2\} & [4.3,3.0] & [.92,.88] &  \\
\hline
0.5 & 1.61 & (21,34) & (25,31) &  \\
 & \{0.2\} & [6.0,4.1] & [1.11,1.09]  & \\
\hline
0.4 & 2.02 & (23,30)   & & \\
 & \{0.2\} & [8.7,6.6]   & & \\
\hline
\end{array} 
\] \label{8msun}
\end{table}
It is tempting to identify the observed line profile variability (LPV) in the well observed Be star $\mu$ Cen (Rivinius et al. \cite{riv01}) in terms of $n=-1$ q-mode oscillations in a rapidly rotating B-star.  The observed variations are indeed all {\it prograde} in the observer's frame and fall in two narrow period intervals of about 0.50 days ($m$=2) and 0.28 days ($m$=3). Qualitatively, this is exactly what is found for the unstable $n=$-1 q-modes calculated here (see Table~\ref{8msun}), although the exact periods do not match. But that is not surprising in view of the over-simplistic modelling of extremely rapidly (differentially) rotating Be-stars.

\section{Summary and discussion}
In this paper we study the stability of quasi g-modes ('q-modes'), a  branch of oscillation modes that exists in between the normal retrograde g-modes and the r-modes in rotating stars. It appears that these retrograde (in corotating frame) low frequency oscillations can be destabilized by the $\kappa$-mechanism in the driving region connected with the opacity bump close to the stellar surface. From relatively low angular rotation speeds upwards the unstable q-modes with $m>1$ occur in a few narrow period bands (for different $m$-values) in which the different radial orders lie densely packed together.  For $m$=1 the periods of q-modes are typically larger than a day, up to many days for low radial orders. In the observer's frame all q-modes are prograde. The predicted observable q-mode spectrum is thus in sharp contrast to the predicted spectrum of unstable g-modes which occur in much wider period intervals, whereby modes of different radial order $k$ should be separately observable with even modest time resolution. 
The unstable q-modes seem to fit the observed rather scarce frequency spectrum in SPB stars and $\beta$ Cephei stars (e.g. Stankov et al. \cite{stank02})  better than normal g-modes. The observed closely spaced frequencies in e.g. the SPB star HD 160124 (Waelkens, \cite{waelkens91}) cannot be explained by unstable g-modes, but is typical for the q-mode spectrum. 
It is, however, doubtful whether the rotational confinement of g-modes towards the stellar equator in rotating stars can explain away the many predicted unstable g-modes (see Townsend \cite{town05}).  It remains mysterious why we don't observe the many predicted unstable g-modes and would instead observe the (weaker) unstable q-modes. It helps in any case that the unstable q-modes hardly have a tendency to focus on the equatorial region of rotating stars. 

Although the more massive $\beta$ Cephei pulsators seem to show clustered periods similar to SPB stars (but with smaller periods) they cannot be explained straightforwardly in terms of q-modes because in stars more massive than 8$M_\odot$ the thermal timescale in the the Z-bump driving region becomes too short to enable destabilization of the q-modes. Q-mode instability for these more massive $\beta$ Cephei pulsators would require the Z-bump region to be deeper inside the star than predicted by the current stellar evolution models.

Nevertheless, the observed LPV in Be stars like $\mu$ Cen is (qualitatively) consistent with the calculated properties of q-modes in a rapidly rotating 8M$_\odot$ B-star: two narrow period intervals corresponding to two unstable series (for $m$=2 and $m$=3) of closely spaced q-modes, see previous section. One may speculate about the cause of the related outbursts in this Be-star in terms of q-modes. Excitation of the (negative angular momentum) retrograde q-modes in the star will accelerate the rotation of the stellar material in the driving region just beneath the stellar surface, perhaps bringing the surface layers up to ejection velocities. 
A good understanding of the Be phenomenon  requires a much more detailed study with differential rotation including the centrifugal force and non-linear effects.


\begin{thebibliography}{}
\bibitem[1994]{balona94}
Balona, L. 1994, MNRAS, 267, 1060

\bibitem[1994] {balonaK94}
Balona, L. \& Koen, C. 1994, MNRAS, 267, 1060

\bibitem[1978]{berth78}
Berthomieu, G., Gonczi, G., Graff, Ph., Provost, L. \& Rocca, A. 1978,
A \& A, 70, 597

\bibitem[1996]{bild96}
Bildsten, L., Ushomirsky, G. \& Cutler, C. 1996, ApJ, 460, 827

\bibitem[1941]{cowling41}
Cowling, T.G. 1941, MNRAS, 101, 367

\bibitem[1993]{dziembowski93}
Dziembowski, W.A., Moskalik, P. \& Pamyatnykh, A.A. 1993, MNRAS, 265,588

\bibitem[1930] {eddington}
Eddington, A.S. 1930, `The internal constitution of the
stars', Cambridge University Press

\bibitem[1972]{egg72}
Eggleton, P.P. 1972, MNRAS, 156, 361

\bibitem[1993]{gautschy93}
Gautschy, A. \& Saio, H. 1993, MNRAS, 262, 213

\bibitem[1996]{gautschy96}
Gautschy, A. \& Saio, H. 1996, Ann. Rev. Astron. Astrophys. 34, 551


\bibitem[1996] {iglesias96}
Iglesias, C.A. \& Rogers, F.J. 1996, ApJ, 464, 943

\bibitem[1992]{kiria92}
Kiriakidis, M., El Eid, M.F. \& Glatzel, W. 1992, MNRAS, 255, 1

\bibitem[2001]{lee01}
Lee, U. 2001, ApJ, 557, 311

\bibitem[1997]{lee97}
Lee, U. \& Saio, H. 1997, ApJ, 491, 839

\bibitem[1978]{pap78}
Papaloizou, J.C.B. \& Pringle, J.E. 1978, MNRAS, 182, 423

\bibitem[1997] {pap97}
Papaloizou, J.C.B. \& Savonije, G.J. 1997, MNRAS, 291, 633

\bibitem[1995]{pols95}
Pols, O.R., Tout, C.A. \& Eggleton, P.P. 1995, MNRAS, 274, 964

\bibitem[1992] {press92}
Press, W.H., Teukolsky, S.A., Vetterling, W.T. et al. 1992,
Numerical Recipes in Fortran, Cambridge University Press

\bibitem[2001]{riv01}
Rivinius, Th. et al. 2001, A \& A, 369, 1058

\bibitem[1992]{iglesias92}
Iglesias, C.A., Rogers, F.J. \& Wilson, B.G. 1992, ApJ, 397,717

\bibitem[1983]{sav83}
Savonije, G.J. \& Papaloizou, J.C.B. 1983, MNRAS, 203, 581

\bibitem[1995] {sav95}
Savonije, G.J., Papaloizou, J.C.B. \& Alberts, F. 1995, MNRAS, 277, 471

\bibitem[2002]{sav02}
Savonije, G.J. \& Witte, M.G. 2002, A \& A, 386, 211

\bibitem[2002] {stank02}
Stankov, A., Handler, G., Hempel, M. \& Mittermayer, P. 2002, MNRAS, 336, 189

\bibitem[2003]{town03}
Townsend, R.H.D. 2003, MNRAS, 343, 125

\bibitem[2005]{town05}
Townsend, R.H.D. 2005, astro-ph/0503192.
\bibitem[1989]{unno89}
Unno, W., Osaki, Y., Ando, H., Saio, H. \& Shibahashi, H. 1989,
Non-radial Oscillations of Stars. University of Tokyo Press, Tokyo

\bibitem[1998]{usho98}
Ushomirsky, G. \& Bildsten, L. 1998, ApJ, 497, L101

\bibitem[1991] {waelkens91}
Waelkens, C. 1991, A \& A, 246, 453


\end{thebibliography}
\end{document}